\documentclass[a4paper,11pt]{article}
\pdfoutput=1 

\usepackage{jheppub} 
\usepackage[dvipsnames]{xcolor}
\usepackage[T1]{fontenc} 
\usepackage{CJKutf8}
\usepackage{mathrsfs}
\usepackage{tikz}
\usetikzlibrary{decorations.pathmorphing}
\usetikzlibrary{decorations.markings}

\newcommand{\lrnabla}{\overleftrightarrow\nabla}
\newcommand{\moe}{\sigma}
\newcommand{\CDv}{\mathcal K}
\newcommand{\tCDv}{\widetilde{\mathcal K}}
\newcommand{\tcU}{\widetilde{\mathcal U}}
\newcommand{\Prp}{\Delta}
\newcommand{\mqInt}{\int \!d\bar Q}
\title{\boldmath Covariant derivative expansion for the renormalization of gravity}

\author{ Rodrigo Alonso}

\affiliation{Kavli Institute for the Physics and Mathematics of the Universe (WPI), The University of Tokyo Institutes for Advanced Study, The University of Tokyo, Kashiwa, Chiba 277-8583, Japan}

\emailAdd{rodrigo.alonso@ipmu.jp, rodrigo.alonso@cern.ch}

\abstract{The one loop UV divergences of Hilbert-Einstein gravity with a cosmological constant and spin 0, 1/2 and 1 matter are computed  making use of a covariant derivative expansion and functional methods. For this purpose the transformation that yields the covariant derivative~\cite{Gaillard:1985uh} is extended to include a dynamical metric and the expansion in the fields themselves is made covariant which is relevant for the effective action due to the non-linear character of gravity.}

\begin{document} 
	\begin{flushright}
IPMU 19-0183
	\end{flushright}
\maketitle

\section{Introduction}
\label{sec:intro}
The last decade has seen the rise of effective field theory (EFT) to the forefront of particle physics as the mainstream general framework to process experimental data into theory. This shift was originated by experimental data and the absence of long-heralded evidence in it but at the same time EFT brings changes to the theorist perspective also. The non-so-aptly named non-renormalizable theories possess a well defined and computable pertubative expansion with a finite set of parameters at any given order in couplings and loop expansion. Indeed the recent surge in activity has produced quantum level general ~\cite{Jenkins:2013zja,Jenkins:2013wua,Alonso:2013hga,Alonso:2014zka,Elias-Miro:2013mua,Henning:2014wua,Drozd:2015rsp}, and new~\cite{Alonso:2014rga,Cheung:2015aba,Bern:2019wie,Henning:2015alf,Henning:2019enq} results and automatization~\cite{Criado:2017khh,Bakshi:2018ics,Celis:2017hod}. While these works are inspired by the reasons that mark the Standard Model (SM) as incomplete, there is another theory of nature which requires completion, gravity, and it does fit the mold of EFT seamlessly~\cite{Donoghue:1994dn,Donoghue:1995cz,Donoghue:2012zc}. Here in an effort to bring the two closer together, techniques developed in the context of the SM EFT will be generalized to dynamical gravity. To be specific, by means of a covariant derivative expansion~\cite{Gaillard:1985uh}, the UV divergences at one loop generated by gravitational interactions for Hilbert-Einstein gravity with a cosmological constant (CC) and scalar, fermions and vector bosons will be computed. A good deal of the final results for UV divergences here obtained have been in the literature for some time~\cite{tHooft:1974toh,Deser:1974cy,Deser:1974cz}, and with the heat-kernel method~\cite{Avramidi:2000bm} general results at the loop level are available~\cite{Fradkin_1977,Christensen:1984dv,Barvinsky:1985an,Vilkovisky:1992pb} by computation of DeWitt coefficients~\cite{DeWitt:1965jb}. The novel aspect of this work is therefore the technique for the computation, which we hope makes the derivation of results in quantum gravity more accessible for a particle physicist while the quantum gravity practitioner might find the reduced mathematical machinery makes some aspects of the quantum structure of gravity more pristine.

Section~\ref{Del2S} lays out the functional formulation of one loop corrections and computes the field-covariant second order variation of the action while sec.~\ref{CDEGR} presents the transformation and the resulting covariant derivative for gravity. Sec.~\ref{TrLogEv} combines the previous results to compute the UV divergences at one loop.
Our conventions are a flat metric as $\eta_{\mu\nu}=$Diag$(1,-1,-1,-1)$ and 
\begin{align}
\nabla_\mu A^{\alpha}&=\partial_\mu A^\alpha+\Gamma_{\mu\nu}^\alpha A^\nu&  [\nabla_\mu,\nabla_\nu] A^{\alpha}&\equiv  R^{\alpha}_{\,\,\,\beta\mu\nu}A^\beta &R_{\mu\nu}&\equiv R^{\alpha}_{\,\,\,\mu\alpha\nu}
\end{align}
where we note that part of the literature uses an opposite-sign definition for $R_{\mu\nu}$~\cite{Donoghue:1994dn}. Given that in sec.~\ref{TrLogEv} dimensional regularization is used we write our formulae in $d$ dimensions with $d$ in the vicinity of 4.

\section{Second order covariant variation of the action\label{Del2S}}
Functional methods have been applied to particle physics over the decades and the recent literature contains complete and accessible descriptions~\cite{Henning:2014wua,Drozd:2015rsp} to which we refer the reader for the detailed formulation; here rather we shall start from a number of results in the literature whose combination is required to tackle gravity. The one-loop corrections to the action can be synthesized into a Gaussian integral as, formally, 
\begin{align}
 e^{iS[\hat\phi]_{\rm eff}}=\int D\delta \phi e^{iS[\hat \phi]+i\delta \phi \delta S[\hat \phi]+\frac i2 (\delta \phi)^2\delta^2S[\hat\phi]+\mathcal O(\delta\phi^3)}\simeq e^{iS[\hat\phi]-\frac12 {\rm tr}({\rm log}(-\delta^2 S[\hat\phi])) }\,,
\end{align}
with $\hat \phi$ the background field, $S_{\rm eff}$ the effective action and the last equality valid to one loop.
The one point to be underlined here is that, if one were to use a different variable for the field related as $\phi=\phi(\varphi)$ the second variation $\delta^2 S$ does not transform as a true tensor,
\begin{align}
( \delta \phi)^2 \frac{\delta ^2 S}{\delta \phi\delta \phi}=\left(\delta \varphi \frac{\delta \phi}{\delta \varphi}\right)^2 \frac{\delta ^2 S}{\delta \phi\delta \phi}=(\delta \varphi)^2 \frac{\delta ^2 S}{\delta \varphi\delta \varphi} -(\delta \varphi)^2\frac{\delta^2 \phi }{\delta\varphi\delta\varphi}\frac{\delta S}{\delta \phi}\,,
\end{align}
this one can remedy making use of a (true) 2-tensor, the metric in field space:
\begin{align}
\partial_\mu \phi G(\phi)\partial^\mu\phi\to \partial_\mu \varphi \frac{\partial\phi}{\partial\varphi} G(\phi)\frac{\partial\phi}{\partial\varphi}\partial^\mu\varphi=\partial^\mu\varphi G' (\varphi)\partial^\mu\varphi\,,
\end{align}
and a covariant derivative {\it in field space}~\cite{Honerkamp:1971sh} $\mathcal D_i V^j= \delta_i V^j+\hat\Gamma^{j}_{ik} V^k$. In particular for the action (taken to be a scalar) we have:
\begin{align}\label{ConFldS}
\mathcal D S=\frac{\delta S}{\delta \phi}\,,&&\mathcal D^2S=\frac{\delta^2 S }{\delta\phi^i\delta\phi^j }-\hat \Gamma^k_{ij}\frac{\delta S}{\delta \phi^k}\,,&&
\hat\Gamma=\frac{(G^{-1})^{kl}}{2}\left(\frac{\delta G_{li}}{\delta \phi^j}+\frac{\delta G_{jl}}{\delta \phi^i} - \frac{\delta G_{ij}}{\delta \phi^l}\right)\,,
\end{align}
where we note that this applies even if one started with a constant metric $G$ and for some reason wanted to perform a non-linear change of field variable.  In this way the covariant one loop action result, including the invariant measure in field space $\sqrt{G}D\phi$, reads to the one-loop level
\begin{align}
iS_{\rm eff}[\hat \phi]={\rm log}\left(\int\! \!\sqrt{G} D\delta\phi \,e^{iS+i\delta \phi \mathcal D S+i\delta \phi^2\mathcal D^2 S/2}\right)=iS[\hat\phi]-\frac12 {\rm tr}({\rm log}(-(\mathcal D^2 S[\hat\phi]) G^{-1}))\,, \label{Origin}
\end{align}
where the product $(\mathcal D^2 S[\hat\phi]) G^{-1}$ makes an operator with a covariant and a contra-variant index in field-variable-indexes and hence the trace is an `invariant' result, meaning an expression for which physicists who choose to describe a system with different field variables agree on. This covariant description does as well preserve the (linear \& non-linear) symmetries of the original action at the loop level
which one can realise in this formalism as a specific change of variable.

Let us then turn to the action at hand to first determine $(\mathcal D^2 S[\hat\phi]) G^{-1}$,  here considered is the Hilbert-Einstein action with a cosmological constant and spin 0,1/2 and 1 matter,
\begin{align}\label{OrAct}
S=\int dV\left(\frac{1}{2\kappa^2} (2\Lambda-R)+\frac12\left( \nabla_\mu\phi\nabla^\mu\phi-m_\phi^2\phi^2\right)+ \psi^\dagger \moe^\mu
\frac{i\overleftrightarrow{\nabla}_\mu}{2} \psi+ \frac14 F_{\alpha\beta}F^{\beta\alpha}\right)\,,
\end{align}
with $dV=d^dx\sqrt{-g}$, $\kappa^2=8\pi G_N$ where $G_N$ is Newton's constant. This action describes the Standard Model (SM) plus gravity in the limit of vanishing SM couplings (gauge, Yukawa and quartic) and so with $\Lambda\sim 4\times 10^{-66}$eV we believe it describes nature in said limit. For the covariant action  the first variation of the action w.r.t. the metric is needed
\begin{align}\nonumber
\frac{\delta S}{\delta g_{\mu\nu}}=\int dV\Bigg(&-\frac{1}{2\kappa^2}\left(\frac{g^{\mu\nu}}{2}\left(R-2\Lambda \right)-R^{\mu\nu}\right)+\frac12\left(\frac{g^{\mu\nu}}{2}\left(\partial\phi^2-m_\phi^2\phi^2\right)-\partial^\mu\phi \partial^\nu\phi\right)\\&+\frac{i}{4}\psi^\dagger\left(g^{\mu\nu}\moe\overleftrightarrow\nabla-\frac{\moe^{\mu}\overleftrightarrow\nabla^{\nu}+\moe^{\nu}\overleftrightarrow\nabla^{\mu}} 2\right)\psi
+\frac{1}{8}g^{\mu\nu}(F F)-\frac12 (F F)^{\mu\nu}\Bigg)\,,
\end{align}
whereas for matter fields we have linear realizations, that is, with the chosen variables their `metrics' are flat and hence $\hat\Gamma[\phi,\psi,A]=0$.
 The metric itself ($g_{\mu\nu}$) in contrast does have a `metric' ($G^{\mu\nu,\rho\sigma}$), not to dwell in linguistics let us anticipate results and simply give it here:
\begin{align}\label{MetCon}
G^{\alpha\beta,\sigma\rho}(g)=&\frac14\left(g^{\alpha(\sigma}g^{\rho)\beta}-g^{\alpha\beta}g^{\rho\sigma}\right)\,, &&	\hat\Gamma^{\alpha\beta,\rho\sigma}_{\mu\nu}=-\frac18\, g^{(\alpha}_{\,\,(\mu}g^{(\rho}_{\,\,\nu)} g^{\beta)\sigma)}\,,
\end{align}
where parenthesis around indixes denotes symmetrization $V_{(\alpha} W_{\beta)}=V_\alpha W_\beta+V_\beta W_\alpha$ and with the opposite placing of indices as usual  yet this convention follows from our component field $g_{\mu\nu}$. This somewhat unfamiliar language might be more accessible if we note that in the graviton propagator or the `inverse' of the two point action has in it the inverse of the metric $G$, $G^{-1}_{\alpha\beta,\rho\sigma}=g_{\alpha(\sigma}g_{\rho)\beta}-g_{\alpha\beta}g_{\rho\sigma}$. Otherwise this treatment for a covariant result is not new in gravity and is related to what is at times termed a Vilkovisky's action~\cite{Vilkovisky:1984st}.

The covariant second order variation then reads
\begin{align}
\mathcal D^2S\equiv\frac12\delta g^2\mathcal D^2S+\frac12\delta\Phi^2\frac{\delta^2S}{\delta\Phi\delta\Phi}=\frac12\left((\delta g\frac{\delta^2 S}{\delta g\delta g}\delta g)+(\delta g\frac{\delta S}{\delta g}\delta g)\right)+\frac12\delta\Phi^2\frac{\delta^2S}{\delta\Phi\delta\Phi}\,.
\end{align}
Next the explicit expression for $(\mathcal D^2 S[\hat\phi]) G^{-1}$ arising from each piece of the action in~(\ref{OrAct}) is given, for which purpose we define:
\begin{align}
&S^{(2)}_{n}=\frac12\delta \phi^2 \mathcal D^2 S_{n}=\int dV \mathscr{L}^{(2)}_{n}\,,& &\{S_n\}=\{S_g\,,\,S_\phi\,,\,S_\psi\,,\,S_{A}\}\,.
\end{align}

\subsection{Hilbert-Einstein and cosmological constant}
The covariant second order variation of the Hilbert-Einstein action with a cosmological constant reads (with an abuse of notation we compute variations from eq.~(\ref{OrAct}) with $g_{\mu\nu}\to g_{\mu\nu}+\delta g_{\mu\nu}$ so that the background field is $g$ which is also understood to raise and lower indices from now on)
\begin{align}\nonumber
S^{(2)}_g=\int \frac{-\sqrt{|g|}}{4\kappa^2}\Big(&(\delta g) \nabla^\alpha\nabla^\beta \delta g_{\alpha\beta}-\delta g_{\alpha\beta} \nabla^\beta\nabla^\rho\delta g_{\rho}^{\,\,\alpha}+\frac12\delta g_{\alpha\beta}\nabla^2\delta g^{\alpha\beta}-\frac{1}{2}(\delta g)\nabla^2(\delta g)\\
&+R^{\alpha\rho\beta\sigma}\delta g_{\alpha\beta}\delta g_{\rho\sigma}-(\delta g) R^{\alpha\beta}\delta g_{\alpha\beta}+\frac{R-2\Lambda}{4}(\delta g)^2
\Big)d^dx\,,
\end{align}
where a two-index object within parenthesis means it is traced over, $(\delta g)=\delta g_{\mu\nu}g^{\mu\nu}$. As with other gauge theories, the path integral has a large redundant integration volume associated here to the linearised symmetry:
\begin{align}
\delta g_{\epsilon}=\delta g_{\mu\nu}+\nabla_{(\nu} \epsilon_{\mu)}\,,
\end{align}  which one disposes of with the Faddeev-Popov procedure. The function , $\mathcal X_\mu(\delta g)=\nabla^. \delta g_{.\mu}-\nabla_\mu(\delta g)_\mu$ is used for gauge fixing and requires of an extra term in the action
\begin{align}
1&=\int\! D\epsilon \delta \left(\mathcal X(g_\epsilon)\right)
{\rm det} \left(\frac{\delta \mathcal X(\delta g_{\epsilon})} {\delta \epsilon^\mu}\right)=
\int \!D\epsilon \delta \left(\mathcal X(g_\epsilon)\right)\int\! D\bar c D c	e^{-i\int dV \bar c^\mu\left(g_{\mu\nu}\nabla^2+R_{\mu\nu}\right)c^\nu}\,,
\end{align}
with $c_\mu$ the wrong-statistics auxiliary field, our ghosts, and  adding the term
\begin{align}
	S_\xi=\int\frac{1}{8\kappa^2\xi}\left(\nabla^\nu \delta g_{\nu\mu}-\frac12\nabla_\mu(\delta g)\right)^2dV\,,
\end{align}
leads to the Harmonic gauge when $\xi=1$ which is selected here for computational simplicity. In this gauge the kinetic term reads:
\begin{align}
-\frac{1}{4\kappa^2}\left(\frac{\delta g_{\mu\nu}}{2}\nabla^2\delta g_{\mu\nu}-\frac 14 (\delta g) \nabla^2(\delta g)\right)=-\frac{\delta g_{\alpha\beta}}{4\kappa^2}\nabla^2\left(\frac14 g^{\alpha(\rho}g^{\sigma)\beta}-\frac14g^{\alpha\beta}g^{\rho\sigma}\right)\delta g_{\rho\sigma}\,,
\end{align}
from where the metric in eq.~(\ref{MetCon}) follows.  Note that as for the overall normalization this metric yields off-diagonal components as $\delta g G\delta g= \delta g_{i<j}^2+...$ for a flat metric.
As a final step we raise the index of one of the variations with the metric $G$ so that the resulting operator is ready to be traced over which results in a remarkably simple expression:
\begin{align}
S_{g+\xi+c}^{(2)}=&-\int dV \bar c^\mu\left(g_{\mu\nu}\nabla^2+R_{\mu\nu}\right)c^\nu\\\nonumber
&- \int \frac{1}{4\kappa^2} \delta g_{\alpha\beta}\left(g^{\alpha}_{\,\,(\rho}g^{\beta}_{\,\,\sigma)}\frac{\nabla^2}{2}+R^{\alpha\,\,\,\,\,\beta}_{\,\,\,(\rho\,\,\,\sigma)}-g^{\alpha\beta} R_{\sigma\rho}+\Lambda g^{\alpha\beta}g_{\rho\sigma}\right)(G\cdot\delta g)^{\rho\sigma}dV\,.
\end{align}

\subsection{Scalars}
The addition of a scalar field brings an extra contribution to the graviton variation as well as mixed $\phi-g$ terms:
\begin{align}\nonumber
S^{(2)}_\phi=\int \Big(-&\frac12\delta \phi \nabla^2\delta \phi+\frac14\left((\partial \phi\delta g \delta g\partial\phi)-(\delta g)(\partial\phi\delta g\partial \phi)+\frac14 (\delta g)^2((\partial\phi)^2-m_\phi^2\phi^2)  \right)\\
-&(\partial\phi\delta g\partial\delta \phi )+\frac{(\delta g)}{2}(\partial\phi\partial\delta \phi-m_\phi^2\phi\delta\phi)\Big)dV\,,
\end{align}
where again a two-index object within parenthesis means it is traced over and $\delta g$ in between $\partial\phi$ are taken as vector-matrix scalar products, e.g $(\partial\phi \delta g \partial\phi)=\partial^\mu\phi\delta g_{\mu\nu}\partial^\nu\phi$.
The mixed terms are removed here completing squares without modifying the measure~\cite{Henning:2016lyp}:
\begin{align}
\delta \phi\to \delta \phi-\frac{1}{\nabla^2+m_\phi^2}\left(\frac{(\nabla \partial\phi (\delta g))+m_\phi^2\phi}{2}-(\nabla \delta g \partial \phi)\right)\,.
\end{align}
This results into, after raising the index in the graviton variation
\begin{align}\label{ScUg}
 \mathscr{L}^{(2)}_\phi=& -\frac12\delta \phi (\nabla^2+m_\phi^2)\delta\phi\\ \nonumber
&-\frac{\delta g_{\alpha\beta}}{4\kappa^2}\left(\kappa^2g^{\alpha\beta}\left(\phi_{;\rho}\phi_{;\sigma}-\frac{g_{\rho\sigma}(m_\phi\phi)^2}{2}\right)-\frac{\kappa^2}{2}\phi^{;(\alpha}\phi_{;(\rho}\, g^{\beta)}_{\,\,\,\sigma)} \right)(G\cdot \delta g)^{\rho\sigma}\\ \nonumber
&-\frac{\delta g_{\alpha\beta}}{4\kappa^2}\left(\left(g^{\mu(\alpha}\phi^{;,\beta)}-g^{\alpha\beta}\phi^{;\mu}\right)\nabla_\mu+m_\phi^2\phi g^{\alpha\beta}\right)\frac{\kappa^2}{\nabla^2+m_\phi^2}\left(\nabla_{(\rho} \phi_{;\sigma)}+g_{\rho\sigma}m_\phi^2\phi \right) (G\delta g)^{\rho\sigma}
\end{align}
where, to keep the equations of manageable length we have used the semi-colon notation $\phi_{;\alpha}=\nabla_{\alpha}\phi$  and the explicit $\nabla$'s are to be taken as acting on everything on their right, termed `open' derivatives. 

A global transformation as $g_{\mu\nu}\to(1+ \alpha) g_{\mu\nu}$, $\delta \phi \to(1+ \frac{2-d}{4}\alpha )\delta \phi$ leaves the action the same (for $m_{\phi}\to0$) whereas one can change the scalar action into
\begin{align}
\mathscr{L}_{\phi_{CFT}}=-\frac12\phi\left(\nabla^2-\frac{d-2}{4(d-1)}R\right)\phi\,,
\end{align}
for a locally scale-invariant action.

\subsection{Fermions}
The diffeomorphism-invariant Weyl-fermion kinetic term in eq.~(\ref{OrAct}) is, explicitly
\begin{align}
\frac i2 \psi^\dagger \sigma_\mu\lrnabla\psi
=\frac i2 \psi^\dagger \sigma^c e^\mu_c\left(\partial_\mu +\frac{\bar\sigma^{[a}\sigma^{b]}}{8}e^a_{\nu}(\partial_\mu e^{b,\nu}+\Gamma_{\mu\rho}^\nu e^{b,\rho} ) \right)\psi+h.c.
\end{align}
where $e^{\mu}_ae_b^{\nu}\eta^{ab}=g^{\mu\nu}$, $\sigma^a=(1,\vec\sigma)$, $\bar\sigma^a=(1,-\vec\sigma)$, and $\psi$ is a RH fermion ($\psi^{\dot\alpha}$). 
In the following a Greek letter (or symbol) as index for the sigma matrices denotes contraction with the vierbein $\sigma\cdot e_\mu=\sigma_ae^a_\mu\equiv\moe_{\mu}$. 

The second order covariant action is
\begin{align}\nonumber
S^{(2)}_\psi=\int\frac {i}{2}&\Big[\,\delta \psi^\dagger \moe \nabla \delta \psi-h.c.+\frac{i(\nabla_\mu\delta g_{\alpha\beta})\delta g^\beta_{\,\,\,\rho}}{8}\psi^\dagger \varepsilon^{\mu\alpha\rho\nu}\moe_{\nu}\psi\\ \nonumber
&+\left( \frac{(\delta g)^2}{8}\psi^\dagger \moe\nabla\psi+\frac 18 \psi^\dagger \moe \delta g\delta g\nabla \psi-\frac{\delta g}{4}\psi^\dagger \moe\delta g\nabla \psi\right)-h.c.\\
&+\left(\delta\psi^\dagger\frac{(\delta g)\moe\nabla-(\moe\delta g \nabla)}{2}\psi+\psi^\dagger\frac{(\delta g)\moe\nabla-(\moe\delta g \nabla)}{2}\delta\psi\right)-h.c. \Big]dV\,,
\end{align}
with $\varepsilon^{\mu\nu\rho\lambda}=e_a^\mu e_b^\nu e_c^\rho e_d^\lambda\epsilon^{abcd}$, $\epsilon^{0123}=1$. Here as well a field redefinition of the integrating field $\delta \psi$ can be used as
\begin{align}
\delta \psi\to \delta \psi -\frac{1}{\moe\nabla}\frac{(\delta g)\moe\nabla-(\moe\delta g \nabla)}{2}\psi\,,
\end{align} to reduce the action to diagonal form
\begin{align}\nonumber
\mathscr{L}^{(2)}_\psi=	\frac {i}{2}&\Big[\,\delta \psi^\dagger \moe \nabla \delta \psi-h.c. +\frac{i}{8}(\delta g \nabla_\mu\delta g)_{\rho\alpha}\psi^\dagger \varepsilon^{\mu\alpha\rho\nu}\sigma_\nu\psi\\ \nonumber
	&+\left( \frac{(\delta g)^2}{8}\psi^\dagger \moe\nabla\psi+\frac 18 \psi^\dagger \moe \delta g\delta g\nabla \psi-\frac{(\delta g)}{4}\psi^\dagger \moe\delta g\nabla \psi\right)-h.c.\\
	&-\left(\frac{1}{\moe\nabla}\frac{(\delta g)\moe\nabla-(\moe\delta g \nabla)}{2}\psi\right)^\dagger\frac{(\delta g)\moe\nabla-(\moe\delta g \nabla)}{2}\psi-h.c.\quad\Big]\,,
\end{align}
this variation, modulo the equation of motion piece, agrees with the Feynman rule for a two-graviton two-fermion vertex as in~\cite{Bjerrum-Bohr:2014lea}.
The raising of the rear index of the operator in metric space reads
\begin{align}\label{S2psi}
\mathscr L^{(2)}_\psi=\frac{i}{2}\delta \psi^\dagger\moe&\lrnabla\delta\psi\\ \nonumber -\frac{\delta g_{\alpha\beta}}{4}\Bigg[&\frac{g_{\rho\sigma}}{4}\left(g^{\alpha\beta}\psi^\dagger i\moe_\mu\psi^{;\mu}-\frac{\psi^\dagger i\moe^{(\alpha} \psi^{;\beta)}}{2}\right)+h.c.-\frac{1}{16}\left\{\psi^\dagger\varepsilon^{\mu(\alpha\,\,\,\,\nu}_{\,\,\,\,\,\,\,\,(\rho}\sigma_\nu\psi g^{\beta)}_{\sigma)}\psi,\nabla_\mu\right\}\\ \nonumber&
+\frac{g^{\alpha\beta}\psi^\dagger i\moe_{(\rho} \psi_{;\sigma)} }{4}
-\frac{g^{(\beta}_{(\sigma}\psi^\dagger\left(i\moe^{\alpha)} \psi_{;\rho)}+i\moe_{\rho)}\psi^{;\alpha)}\right)}{16}+h.c.\\ \nonumber
& +\frac12\left((\psi^{;\mu})^\dagger\moe_\mu g^{\alpha\beta}-\frac{(\psi^{;(\alpha})^\dagger\moe^{\beta)}}{2}\right)\frac{i}{\moe\lrnabla}\left(g_{\rho\sigma}\moe^\nu\psi_{;\nu}+\moe_{(\rho}\psi_{;\sigma)}\right)\Bigg](G\delta g)^{\rho\sigma}
\end{align}
where once more we resorted to semicolon for derivatives on background fields whereas the remaining $\nabla$ act on anything on its arrow direction and $\{,\}$ is the anticommutator.
Here as in the scalar case one has derivatives acting on the field variation, i.e. `open' derivatives, but as opposed to the spin 0,1 case the action is linear in $\nabla$ which is of relevance for the loop integral analysis as shown in sec.~\ref{CDEGR}. 
In addition we convert the Grassmanian gaussian integral into an opposite-sign scalar integral as $e^{{\rm tr\,log}\mathcal O}= e^{1/2{\rm tr\,log}(\mathcal O\mathcal O^\dagger)}$  for which purpose the following relations are used
\begin{align}
&\nabla _{[\mu}\nabla_{\nu]}\psi=\frac{\sigma^{[a}\bar\sigma^{b]}}8e_{a,\rho}e^\lambda_{b} R^{\rho}_{\,\,\lambda\mu\nu}\psi\,, &&
\moe^\mu\moe^\nu\nabla_\mu\nabla_\nu=\nabla^2-\frac R4\,.
\end{align}

\subsection{Vector boson}
For gauge vector bosons one has a kinetic term, in our matrix notation
\begin{align}
S_{A}=-\int d^dx\frac{\sqrt{-g}}{4}F_{\mu\nu}F_{\alpha\beta}g^{\mu\alpha}g^{\nu\beta}=\int d^dx\frac{\sqrt{-g}}{4} (F\,F)\,,
\end{align}
whose second order covariant variation reads
\begin{align}\nonumber
	S^{(2)}_A=\int dV\Bigg(&\frac14\left( \frac{(\delta g)^2}8(F F)+(F\delta g\delta g F)+(F\delta g F\delta g)-(\delta g) (F\delta g F )\right)\\
	&+\frac14\left((\delta F \delta F)-2(F\delta F\delta g )-2(F\delta g\delta F  )+(\delta g)( F \delta F  )\right)\Bigg)\,.
\end{align}
The gauge symmetry acting on the variation of the vector boson field $\delta A_\mu$ is, in the limit of vanishing gauge coupling,
\begin{align}
(\delta A_\epsilon)_\mu=\delta A_\mu+\nabla_\mu \epsilon(x)\,.
\end{align}
The second order variation on gauge fields, explicitly, is
\begin{align}\nonumber
-&\frac{\sqrt{-g}}2\delta A_\lambda \left(g^{\lambda\alpha}g^{\sigma \beta} \nabla_\beta\nabla_\alpha-g^{\lambda\sigma}\nabla^2\right)\delta A_\sigma\\
=&-\frac{\sqrt{-g}}2\delta A_\lambda \left(g^{\lambda\alpha}g^{\sigma \beta} \nabla_\alpha\nabla_\beta+R^{\sigma\lambda}-g^{\lambda\sigma}\nabla^2\right)\delta A_\sigma\,,
\end{align}
which we supplement with gauge fixing via the function $\mathcal X(\delta A)=\nabla_\mu \delta A^\mu$. The ghost action is not innocuous even for a $U(1)$ symmetry since it involves a  field-dependent ghost Lagrangian as,
\begin{align}
1&=\int D\epsilon \delta \left(\mathcal X( \delta A_\epsilon)\right)
{\rm det} \left(\frac{\delta \mathcal X( \delta A_\epsilon)} {\delta \epsilon}\right)=
\int D\epsilon \delta \left(\mathcal X( \delta A_\epsilon)\right)\int Dc D\bar c	e^{-i\int dV \bar c\nabla^2c}\,,
\end{align}
The gauge fixing term $\mathscr L_\xi=-(\nabla \delta A)^2/(2\xi)$ is added to the action and the Feynman gauge is selected in the following again for computational simplicity. As for the mixed terms, the redefinition that eliminates them  is
\begin{align}
\delta A\to  \delta A_\lambda-\frac12 (\nabla^2-R)^{-1}_{\lambda\omega}\nabla_\mu\left((\delta g F+F\delta g)^{[\omega\mu]}-(\delta g)F^{\omega \mu}\right)\,,
\end{align}
which leaves behind the term
\begin{align}\nonumber
\mathscr{L}_A^{(2)}\supset-\frac{1}{8}\nabla_\mu((\delta g F+ F\delta g)^{[\lambda\mu] })-(\delta g)F^{\lambda\mu})(\nabla^2-R)^{-1}_{\lambda\omega}\nabla_\nu((\delta g F+ F\delta g)^{[\omega\nu]})-(\delta g)F^{\omega\nu})\,,
\end{align}
that combines with the remaining terms to give
\begin{align}\label{S2Amu}
\mathscr L^{(2)}_{A+\xi+c}=\frac{1}{2}
\delta A_\rho&
\left(g^{\rho\sigma}\nabla^2-R^{\rho\sigma}\right)\delta A_\sigma-\bar c \nabla^2 c\\ \nonumber
-\frac{\delta g_{\alpha\beta}}{4\kappa^2}&\Bigg[g_{\rho\sigma}\left((FF)^{\alpha\beta}-\frac{g^{\alpha\beta}}{4}(FF)\right)+g^{\alpha\beta}(FF)_{\rho\sigma}-F^{\alpha}_{\,\,\,(\rho} F^{\,\,\,\,\beta}_{\sigma)}-\frac{(FF)^{(\alpha}_{(\rho} g^{\beta)}_{\sigma)}}{2}\\ \nonumber
&-\left(g^{[\lambda(\alpha}F^{\beta) \mu]}-g^{\alpha\beta}F^{\lambda\mu}\right)\nabla_\mu(\nabla^2-R)^{-1}_{\lambda\omega}\nabla_\nu\left(g^{[\omega}_{(\rho} F_{\sigma)}^{\,\,\nu]}-g_{\rho\sigma}F^{\omega\nu}\right)\Bigg](G\delta g)^{\rho\sigma}\,.
\end{align}

{\bf Collection of formulae}
	
The one loop action then is the sum of the tr log of the operators above as
\begin{align}\nonumber
S_{1\rm loop} =\frac i2\mbox{tr}\left[\log \mathcal O_{\delta g}\right]-i\mbox{tr}\left[\log \mathcal O_{c^\mu}\right]+\frac  i2\mbox{tr}\left[\log \mathcal O_\phi\right]-\frac i 2\mbox{tr}\left[\log \mathcal O_\psi\right]+\frac i2\mbox{tr}\left[\log \mathcal O_A\right]-i\mbox{tr}\left[\log \mathcal O_c\right]\,,
\end{align}
where the operators are, for the different Lorentz representations considered here, 
\begin{align} \label{SmmVar}
\mathcal O_\phi&=\nabla^2 +m_\phi^2\,,& \mathcal O_c&=\nabla^2\,, & \mathcal O_{\psi}&= \nabla^2-\frac{R}{4}\,, &
\mathcal O_A&=g_{\mu\nu}\nabla^2-R_{\mu\nu}\,,
\end{align}
\begin{align}
\mathcal O_{c_\mu}&= g_{\mu\nu}\nabla^2+R_{\mu\nu}\,,& \mathcal O_g&= \frac{g^\alpha_{(\rho} g^\beta_{\sigma)}}{2}\nabla^2
+R^{\alpha\,\,\,\,\,\beta}_{\,\,\,(\rho\,\,\,\sigma)}-g^{\alpha\beta} R_{\sigma\rho}+\Lambda g^{\alpha\beta}g_{\rho\sigma}+\mathcal O_T\,,
\end{align}
where the matter-field-dependent operator $\mathcal O_T$ can be written as
\begin{align}\nonumber
\mathcal O_T\cdot G=&\frac{-2\kappa^2}{\sqrt{|g|}}\left(\mathcal D^2(\sqrt{|g|} \mathscr L_T)+ \mathcal D\left(\frac{\delta \sqrt{|g|} \mathscr L_T }{ \delta \Phi}\right)\frac{1}{\sqrt{|g|}\mathcal O_\Phi}\mathcal D\left(\frac{\delta(\sqrt{|g|} \mathscr L_T)}{\delta \Phi}\right)\right)\\
=&\frac{\kappa^2}{\sqrt{|g|}} \mathcal D (\sqrt{|g|}T)-\frac{\kappa^2}{2}\frac{\delta T}{\delta\Phi}\frac{1}{\mathcal O_\Phi} \frac{\delta T}{\delta \Phi}
\end{align}
where $\mathscr L_T$ is the matter Lagrangian, $T$ is the stress-energy tensor, $-\sqrt{|g|}T=2\delta (\sqrt{|g|}\mathscr L_T)=\mathcal D (\sqrt{|g|}\mathscr L_T)$ and $\mathcal D$ the covariant derivative in metric-field space. The first term above contains the connection $\hat \Gamma$ as in eq.~(\ref{MetCon}) whereas the second term does not since it is made up of first derivatives only. The explicit form of $\mathcal O_T$ here is collected from eqs.~(\ref{ScUg},\ref{S2psi},\ref{S2Amu}).

\section{Covariant derivative transformation \label{CDEGR}}
All the operators obtained from the second order variation of the action have the structure
\begin{align}
\mathcal O_\Phi\equiv\mathbb {I}_{\Phi}\nabla^2+\{\nabla,V\}+U_\Phi(\nabla,x)\,,\label{OpDef}
\end{align}
with the `identity' $\mathbb {I}_\Phi$ being on whatever state we are considering both on Lorentz representation and internal space and $U$ is a series in inverse powers of $\nabla$ starting at degree $0$. To evaluate the tr log of such operator
 one can introduce momentum and position eigenstates  as customary~\cite{Henning:2014wua} and write
\begin{align}
\frac i2 {\rm tr}({\rm log}(\mathcal O))=\frac i2\int d^dx\frac{ d^dq}{(2\pi)^d}\mbox{tr}(e^{iqx}({\rm log}\mathcal O)e^{-iqx})\,,
\end{align}
which specifically turns open derivatives into $ e^{-iqx} \nabla e^{iqx}=iq+\nabla$ where $q$ is taken to be covariant $q_\mu$ as opposed to the contravariant $x^\mu$ so that $d^dqd^dx$ is invariant. This representation turns spacetime derivatives $\partial_\mu$ acting on the `quantum' field one is integrating (tracing) over into $iq$ yet this is not a covariant description; in the present case there is in addition the connection  $\Gamma$ in our covariant derivatives. 
A general and simple way of evaluating the operator in a covariant manner all throughout  is to perform a unitary transformation which turns covariant derivatives into field strenghts, i.e. commutators of $\nabla$~\cite{Gaillard:1985uh}.
The naive application of this procedure to gravity nonetheless does not yield the desired outcome,
\begin{align}
e^{i\partial_q\nabla }e^{-iqx}\nabla_\mu e^{iqx}e^{-i\partial_q\nabla }=e^{i\partial_q\nabla }(iq_\mu+\nabla_\mu) e^{-i\partial_q\nabla }
=iq_\mu+\partial_q^. [\nabla_.,q_\mu]+ \mathcal O(q^{-1})\,,
\end{align}
where $\partial_q=\partial/\partial q_\mu$, $\partial_q\nabla=\partial_q^\mu\nabla_\mu$ and $[\nabla_\mu,q_\nu]$ is $-\Gamma_{\mu\nu}^{\rho}q_\rho$. In addition  this same non-commutativity means that the transformation as in the above is not unitary since:
\begin{align}\label{nonUn}
\left(\partial_q\nabla\right)^\dagger=\overleftarrow{\nabla}\overleftarrow{\partial_q}=\nabla\partial_q =\partial_q\nabla+[\nabla,\partial_q]\,.
\end{align}
The transformation to yield a covariant description must therefore be extended, let us write a transformation $e^{iT}$ and expansion in $q$ as
\begin{align}
&e^{iT}; &T&=\sum_n T_{(n)}\,,&T_{(n)}(\lambda q)&=\lambda^{-n}T_{(n)}(q)\,,
\end{align}
and so using the Baker-Campbell-Hausdorff formula one can expand the matrix product into a sum of nested commutators; for the first few terms
\begin{align}
e^{iT }e^{-iqx}\nabla_\mu e^{iqx}e^{-iT }=e^{iT }\left(iq+\nabla_\mu \right)e^{-iT }=iq_\mu-[T_{(1)},q_\mu]+\nabla_\mu +\mathcal O(q^{-1})\,,
\end{align}
and to first order
\begin{align}
T_{(1)}=\frac12\{\partial_q^\mu\,, \nabla_\mu \} +\frac{1}{4}\{[\partial_q\nabla ,\partial_q^\nu], q_\nu\}\,,
\end{align}
returns  $e^{iT}(iq+\nabla) e^{-iT}=iq+\mathcal O(q^{-1})$. As in the case without gravity the field strength appear, at order $q^{-1}$ which, reads
\begin{align}
	e^{iT }\left(iq+\nabla_\mu \right)e^{-iT }=iq_\mu-[T_{(2)},q_\mu]-\frac12[T_{(1)},[T_{(1)},iq]]+i[T_{(1)},\nabla_\mu]+\mathcal O(q^{-2})\,.
\end{align}
Here in contrast to the flat case and once more due to the non-commutativity of $\nabla$ and $q\,\&\,\partial_q$ one has that terms like $\{[\partial_q^\nu,\nabla_\mu],\nabla_\nu\}/2\subset[T_1,\nabla]$ with open derivatives together with non covariant $\Gamma$ terms appear. This is what complicates the procedure and means one has to iterate and determine $T_{(2)}$ by canceling these terms. Solving for $T_{(2)}$ results in
\begin{align}
	T_{(2)}=&-\frac i8 \{[\partial_q\nabla,\partial_q^\mu],\nabla_\mu\}-\frac{i}{24}\{\left[\partial_q\nabla,[\partial_q\nabla,\partial_q^\mu]\right], q_\mu\}\,,
\end{align}
and 
\begin{align}
		e^{iT }\left(iq+\nabla_\mu \right)e^{-iT }=iq_\mu+\frac i 4\{\partial_q^\nu,[\nabla_\nu,\nabla_\mu]\}+\frac i{12}R^\nu_{\,\,..\mu}\{\partial_q^{.2},q_\nu\}+\mathcal O(q^{-2})\,.
\end{align}
After solving for $T_{(2)}$ nonetheless the order $q^{-2}$ transformed covariant derivative presents still open derivative and non-covariant terms and one iterates the procedure to solve for $T_{(3)}$.
An all-order solution for this transformation could not be found here so the pertinent question is then how many orders in $q^{-1}$ are required to encompass UV divergences which are subject of study of this work; anticipating results from sec.~\ref{TrLogEv}, the answer, for four dimensions, is two more terms,
\begin{align}
T_{(3)}=&-\frac{1}{24}\{[\partial_q\nabla,\partial_q^\mu][\nabla_\mu,\partial_q^\nu] ,\nabla_\nu\} \\\nonumber&-\frac{1}{48}\{[\partial_q \nabla,\partial_q^\mu]\partial_q^\nu,[\nabla_\mu,\nabla_\nu] \}-\frac{1}{48}\{[\partial_q\nabla,\partial_q^\mu][\nabla_\mu,[\partial_q\nabla,\partial_q^\nu]],q_\nu\}+\mathcal O([\nabla,\partial_q]^2)\\ 
T_{(4)}=&-\frac{i}{288} \{\left[\partial_q\nabla,\left[\partial_q\nabla,\left[\partial_q\nabla,\partial_q^\mu\right]\right]\right],\nabla_\mu\}
+\frac{i}{144}\{\left[\partial_q\nabla,\left[\partial_q\nabla,\partial_q^\mu\right]\partial_q^\nu,[\nabla_\mu,\nabla_\nu]\right]\}\\ \nonumber
&-\frac{i}{1440} \{\left[\partial_q\nabla,\left[\partial_q\nabla,\left[\partial_q\nabla,\left[\partial_q\nabla,\partial_q^\mu\right]\right]\right]\right],q_\mu\}+\frac{i}{240}
\{\left[\partial_q\nabla,\left[\partial_q\nabla,\partial_q^\mu\right]\right]\left[\nabla_\mu,\left[\partial_q\nabla,\partial_q^\nu]\right]\right],q_\nu\}\\
&-\frac{i}{1440}\{\left[\partial_q\nabla,\left[\partial_q\nabla,\partial_q^\mu\right]\right]\left[\partial_q\nabla,\left[\nabla_\mu,\partial_q^\nu]\right]\right],q_\nu\}+\mathcal O([\nabla,\partial_q])\nonumber
\end{align}
where by $\mathcal O([\nabla,\partial_q]^n)$ we mean terms which are proportional to the connection $\Gamma$ to the $n$ power (recall $[\nabla,\partial_q]\sim\Gamma\partial_q$) and vanish in an inertial frame $\Gamma\to0$ as opposed to derivative $\partial_x^n\Gamma$ terms. It is rightful to drop the terms we have since the final result for the covariant derivative $e^{iT}(iq+\nabla)e^{-iT}$ will be covariant and given the order we are working at, e.g. we need to consider $[T_{(3)},\nabla]$ so orders $\mathcal O([\nabla,\partial_q])$ must be retained in $T_{(3)}$ but $\mathcal O([\nabla,\partial_q]^2)$ can be dropped as we do.
If one however were to descend one more order these omitted terms will be needed. 

The transformation, to this order, turns the derivative $iq+\nabla$ into:
\begin{align}\label{FistCDE}
e^{iT}(iq_\mu+\nabla_\mu)e^{-iT}=& iq_\mu+\frac i 4\{\partial_q^\nu,[\nabla_\nu,\nabla_\mu]\}+\frac i{12}R^\nu_{\,\,..\mu}\{\partial_q^{.2},q_\nu\}\\ \nonumber
&-\frac 16 \{\left[\partial_q\nabla,[\nabla_\nu,\nabla_\mu]\right],\partial_q^\nu\}-\frac1{24}[\nabla_. ,R^{\nu}_{\,\,\,..\mu}]\{\partial_q^{.3},q_\nu\}\\ \nonumber
&-\frac{i}{16}\{\left[\partial_q\nabla,\left[\partial_q\nabla,[\nabla_\nu,\nabla_\mu]\right]\right] ,\partial_q^\nu \}-\frac{i}{80}[\nabla_.,[\nabla_.,R^{\nu}_{\,\,\,..\mu}]]\{\partial_q^{.4},q_\nu\} \\ \nonumber
&+\frac{i}{48}\{R^\nu_{\,\,\,..\mu}\partial_q^{.3},[\nabla_.,\nabla_\nu]\}+\frac{7i}{720}R^{\nu}_{\,\,\,..\rho}R^{\rho}_{\,\,\,..\mu}\{\partial_q^{.4},q_\nu\}+\mathcal O(q^{-4})\\ \nonumber
&\equiv i(q_\mu+\CDv_{\mu})
\end{align}
where given that $(\partial_q)^n$ is symmetric on its $n$ indices and for brevity we collapse them into `$.$' e.g. $R_{\alpha\beta}\partial_q^\alpha\partial_q^\beta=$ $R_{..}\partial_q^{.2}$ and we defined the `gravitational' covariant derivative $\CDv$.
 Obtaining this transformation is somewhat involved but the process has built-in consistency checks. The term $T_{(i)}$ first enters $e^{iT}(iq+\nabla)e^{-iT}$ at order $i-1$ through $-[T_{(i)},q]$ and it is determined by cancellation of open derivative and non-covariant terms produced by lower order terms, e.g. $[T_{(i-1)},\nabla]$. One has that the number of open derivative and non-covariant terms to be canceled exceeds the number of possible structures in $[T_{(i)},q]$. The system of equations is over-constrained which allows for checking a solution obtained with some minimal set of equations against the remaining conditions.
The necessity of the anti-commutators $\{,\}$ follows from  requiring a unitary transformation as sketched in eq.~(\ref{nonUn}). 
  
  It is useful to organize the expansion in inverse powers of $q$ as with $T$ via the definition:
\begin{align}\label{DefCDvi}
e^{iT}e^{-iqx}(\nabla)e^{iqx}e^{-iT}\equiv &i(q+\CDv) &	\mathcal K&=\sum_n\CDv_{(n)} & \CDv_{(n)}(\lambda q)&= \lambda^{-n}\CDv_{(n)}(q)\\
e^{iT}e^{-iqx}Ue^{iqx}e^{-iT}\equiv &\,\mathcal U&	\mathcal U&=\sum_n\mathcal U_{(n)} & \mathcal U_{(n)}(\lambda q)&= \lambda^{-n}\mathcal U_{(n)}(q)\label{DefcU}
\end{align}
The transformation on a background field function $\hat S(x)$ is, to this order:
\begin{align}
e^{iT}\hat S(x)e^{-iT}=&\hat S+i\partial_q[\nabla,\hat S]-\frac 12 \partial_q^{.2}[\nabla_.,[\nabla_.,\hat S]]-\frac i 6\partial_q^{.3} [\nabla_.,[\nabla_.,[\nabla_.,\hat S]]]+\mathcal O(q^{-4})\\=&
\hat S+i\partial_q^.\hat S_{;.}-\frac 12 \partial_q^{.2}\hat S_{;..}-\frac i 6\partial_q^{.3}\hat S_{;...}+\mathcal O(q^{-4})
\end{align}
with the `$.$' notation for $\partial_q$ of eq.~(\ref{FistCDE}).
It is not always the case however that either $\nabla^2$ or a background field is present, it is sometimes both. Take for instance the following construction that appears on eq.~(\ref{ScUg})
\begin{align}\nonumber
	&e^{iT}\left((iq+\nabla)_{\rho} \phi_{;\sigma}+m_\phi^2g_{\rho\sigma}\phi\right)e^{-iT}=e^{iT}(iq+\nabla)_{\rho}e^{-iT}\, e^{iT}\phi_{;\sigma}e^{-iT}+m_\phi^2g_{\rho\sigma} e^{iT}\phi e^{-iT}\\ \nonumber
	&=\left(iq+i\CDv_{(1)} +\mathcal O (q^{-2})\right)_{\rho}	\left(\phi_{;\sigma}+i\phi_{;\sigma\star}\partial_q^\star+\mathcal O (q^{-2})\right) +m_\phi^2g_{\rho\sigma}(\phi+i\phi_{;\star}\partial_q^\star+\mathcal O (q^{-2}))\\
	&=iq_\rho \phi_{;\sigma}+m_\phi^2g_{\sigma\rho}\phi-q_\rho\phi_{;\sigma\star}\partial_q^\star+\mathcal O(q^{-1})
\end{align}
This is the result for a piece of~(\ref{ScUg}), itself part of the operator $U$ in metric-space.

Last let us address the linear term in derivatives in eq.~(\ref{OpDef}). One has, after the transformation
\begin{align}
e^{iT}e^{-iqx}\mathcal Oe^{iqx}e^{-iT}= -(q+\mathcal K)^2 +i\{\mathcal V,q+\CDv\}+\mathcal U=(iq+i\mathcal K+\mathcal V)^2 +\mathcal U-\mathcal V^2
\end{align}
As in conventional loop integrals a `shift' in our integration variable can remove the linear term only now this `shift' is again a transformation of the operator (note that V is a matrix in whatever spin-space is under consideration). The  transformation $e^{i\mathcal V\partial_q}$ leaves:
\begin{align}\nonumber
e^{i\mathcal V\partial_q}\left(iq+i\mathcal K+\mathcal V\right)e^{-i\mathcal V\partial_q}=&iq-[\mathcal V_\mu,q]\partial_q^\mu+\frac i2[\mathcal V_\mu\partial_q^\mu\,,\mathcal A]+i\CDv_{(1)}+\dots\\
=&iq+i\CDv_{(1)}-[[i\partial_q\nabla,V_\mu] ,q]\partial_q^\mu+\frac i2 [V_\mu\partial_q^\mu,V]+\mathcal O(q^{-2})\\ \nonumber
=&iq+i\CDv_{(1)}+\frac i2 \partial_q^\nu (\nabla_{[\nu}V_{\mu]}+V_{[\nu}V_{\mu]}) -\frac i2\partial_q^\nu \nabla_{(\nu}V_{\mu)}+  O(q^{-2})
\end{align}
Higher order will enter our computation as well but as we shall see their contributions to the UV divergent action cancel and we need not make them explicit here.

The final form of the operator is 
\begin{align}
e^{i\mathcal V\partial_q}e^{iT}e^{-iqx}\mathcal O e^{iqx}e^{-iT}e^{-i\mathcal V\partial_q}\equiv -(q+\tCDv)^2+\tcU
\end{align}
with 
\begin{align} 
e^{i\mathcal V\partial_q}e^{iT}e^{-iqx}(\nabla +V) e^{iqx}e^{-iT}e^{-i\mathcal V\partial_q}&\equiv i(q+\tCDv) \\
 e^{i\mathcal V\partial_q}e^{iT}e^{-iqx}(U-V^2)e^{iqx}e^{-iT}e^{-i\mathcal V\partial_q}&\equiv \tcU \label{DeftcU}
\end{align}
and the action of the full transformation on a background field function is
\begin{align}\nonumber
e^{i\mathcal V} e^{iT}\hat Se^{-iT}e^{-i\mathcal V}=&\hat S+i\partial_q[\nabla,\hat S]-\frac{ \partial_q^2}{2}[\nabla,[\nabla,\hat S]]+\cdots\\&+i\partial_q[\mathcal V,\hat S+i\partial_q[\nabla,\hat S]+\dots]-\frac{\partial_q^2}2 [\mathcal V,[\mathcal V,\hat  S+\dots]]+\dots\\ \nonumber
=&\hat S+i\partial_q [\nabla+V,\hat S]-\frac{\partial_q^2}2 [\nabla,[\nabla,\hat S]]\\
&-\frac{\partial_q^2}2 [V,[V,\hat S]]-\partial_q^2[V,[\nabla,\hat S]]-\partial_q[\partial_q[\nabla,V],\hat S]+\mathcal O(q^{-3})
\end{align}

To close this section the derived transformation is applied to the operators obtained from the second order action of eq.~(\ref{OrAct}) in sec.~\ref{Del2S} to second order in inverse loop momenta.

{\bf Spin $< 2$}\\
The case of lower spin ($< 2$) in this work has a simple operator, in particular all the operators for spin $(< 2)$ have $V=0$ and $U=e^{-iqx}Ue^{iqx}$ has only the zeroth term in the large momenta expansion as follows
\begin{align} \nonumber
	&{\rm Scalar}&& {\rm CFT\,\,scalar}& & {\rm Weyl\,\,Fermion}&& {\rm Gauge\,\,boson}\\[1pt] \hline \nonumber &&&&&&&\\[-10pt]
	U=\quad & m_\phi^2&&-\frac R6&&-\frac R 4\mathbb \delta^{\dot\beta}_{\,\,\,\dot\alpha}&&-R_{\rho}^{\,\,\,\lambda}
\end{align}
with the ghost $c_\mu$ operator having $U=R_{\mu\nu}$ and the ghost $c$, $U=0$. The expansion of $\mathcal U$ in eq.~(\ref{DefcU}) is then
\begin{align}
\mathcal U_{(0)}&=U\,, & \mathcal U_{(1)}&=iU_{;.}\partial_q^.\,,& \mathcal U_{(2)}&=-\frac12 U_{;..}\partial_q^{.2}\,,
\end{align}
and $\tcU=\mathcal U$.

{\bf Graviton}\\
The case of the graviton has a linear term in $\nabla$ induced in our case by fermions, this is extracted from eq.~(\ref{S2psi}):
\begin{align}
\left(V^\mu\right)^{\alpha\beta}_{\rho\sigma}=-\frac{ \kappa^2} {16}\psi^\dagger\varepsilon^{\mu(\alpha\,\,\,\nu}_{\,\,\,\,\,\,\,\,(\rho}\sigma_{\nu}\psi g^{\beta)}_{\sigma)}
\end{align}
On the other hand $U$ has accommodated in this case the mixed graviton-matter terms produced after completing squares in the second order covariant action. These terms do depend on open derivatives $\nabla$ a fact that can be used to tell them apart through the definition
\begin{align}
&U=U_{\rm s}+U_{\rm mx}& &e^{-iqx}\bar U_{\rm s}e^{iqx}= U_{\rm s}
\end{align}
where with the variation computed in sec.~\ref{Del2S} one has, for the single-species operator
\begin{align}
	\frac{\left[ U_{\rm s}\right]^{\alpha\beta}_{\rho\sigma}}{\kappa^2}=&\kappa^{-2}\left(R^{\alpha\,\,\,\,\,\beta}_{\,\,\,(\rho\,\,\,\sigma)}-g^{\alpha\beta} R_{\sigma\rho}+\Lambda g^{\alpha\beta}g_{\rho\sigma}\right)\\ \nonumber	&g^{\alpha\beta}\left(\phi_{,\rho}\phi_{,\sigma}-\frac{g_{\rho\sigma}m_\phi^2\phi^2}2\right)-\frac{1}{2}\phi^{,(\alpha}\phi_{,(\rho}\, g^{\beta)}_{\,\,\,\sigma)}-\frac{ig^{(\beta}_{(\sigma}\psi^\dagger(\moe^{\alpha)}\psi_{;\rho)} +\moe_{\rho)}\psi^{;\alpha)})}{16}+h.c.\\ \nonumber
	&+\frac{g_{\rho\sigma}}{4}\left(g^{\alpha\beta}\psi^\dagger i\moe^\mu\psi_{;\mu}-\frac{\psi^\dagger i\moe^{(\alpha} \psi^{;\beta)}}{2}\right)
	+\frac{g^{\alpha\beta}\psi^\dagger i \moe_{(\rho} \psi_{;\sigma)}}{4}+h.c.\\ \nonumber
	&+g_{\rho\sigma}\left((FF)^{\alpha\beta}-\frac{g^{\alpha\beta}}{4}(FF)\right)+g^{\alpha\beta}(FF)_{\rho\sigma}-F^{\alpha}_{\,\,\,(\rho} F^{\,\,\,\,\beta}_{\sigma)}-\frac{(FF)^{(\alpha}_{(\rho} g^{\beta)}_{\sigma)}}{2}\,,
	\end{align}
meanwhile the mixed term reads
	\begin{align}
\frac{\left[U_{\rm mx}\right]^{\alpha\beta}_{\rho\sigma}}{\kappa^2}=&
\left(\left(g^{\mu(\alpha}\phi^{;,\beta)}-g^{\alpha\beta}\phi^{;\mu}\right)\nabla_\mu+m_\phi^2\phi g^{\alpha\beta}\right)\frac{1}{\nabla^2+m_\phi^2}\left(\nabla_{(\rho} \phi_{;\sigma)}+g_{\rho\sigma}m_\phi^2\phi \right)\\ \nonumber
& +\frac12\left((\psi^{;\mu})^\dagger\moe_\mu g^{\alpha\beta}-\frac{(\psi^{;(\alpha})^\dagger\moe^{\beta)}}{2}\right)\frac{i}{\moe\lrnabla}\left(g_{\rho\sigma}\moe^\nu\psi_{;\nu}+\moe_{(\rho}\psi_{;\sigma)}\right)\\ \nonumber
	&-\left(g^{[\lambda(\alpha}F^{\beta) \mu]}-g^{\alpha\beta}F^{\lambda\mu}\right)\nabla_\mu(\nabla^2-R)^{-1}_{\lambda\omega}\nabla_\nu\left(g^{[\omega}_{(\rho} F_{\sigma)}^{\,\,\nu]}-g_{\rho\sigma}F^{\omega\nu}\right)\,.
\end{align}
In the notation of sec.~\ref{Del2S}, the open derivatives in $U_{\rm mx}$ are $\nabla$'s  whereas for derivatives acting only on the background fields we have use the semicolon`$;$' notation. After the transformation $e^{iT}$ one has, to second order, for the single-species contribution
\begin{align}\nonumber
\frac{\left[ \mathcal{U}^{\rm s}_{(0)}-V^2\right]^{\alpha\beta}_{\rho\sigma}}{\kappa^2}=&\kappa^{-2}\left(R^{\alpha\,\,\,\,\,\beta}_{\,\,\,(\rho\,\,\,\sigma)}-g^{\alpha\beta} R_{\sigma\rho}+\Lambda g^{\alpha\beta}g_{\rho\sigma}\right)-\frac{\kappa^2}{16^2}\psi^\dagger\varepsilon^{\lambda(\alpha}_{\,\,\,\,(\mu}g^{\beta)}_{\nu)}\psi\, \psi^\dagger\varepsilon^{\omega(\mu}_{\,\,\,\,(\rho}g^{\nu)}_{\sigma)}\psi g_{\lambda\omega} \\ \nonumber	&g^{\alpha\beta}\left(\phi_{,\rho}\phi_{,\sigma}-\frac{g_{\rho\sigma}m_\phi^2\phi^2}2\right)-\frac{1}{2}\phi^{,(\alpha}\phi_{,(\rho}\, g^{\beta)}_{\,\,\,\sigma)}-\frac{ig^{(\beta}_{(\sigma}\psi^\dagger(\moe^{\alpha)}\psi_{;\rho)} +\moe_{\rho)}\psi^{;\alpha)})}{16}+h.c.\\  \label{ctU0}
&+\frac{g_{\rho\sigma}}{4}\left(g^{\alpha\beta}\psi^\dagger i\moe^\mu\psi_{;\mu}-\frac{\psi^\dagger i\moe^{(\alpha} \psi^{;\beta)}}{2}\right)
+\frac{g^{\alpha\beta}\psi^\dagger i \moe_{(\rho} \psi_{;\sigma)}}{4}+h.c.\\ \nonumber
&+g_{\rho\sigma}\left((FF)^{\alpha\beta}-\frac{g^{\alpha\beta}}{4}(FF)\right)+g^{\alpha\beta}(FF)_{\rho\sigma}-F^{\alpha}_{\,\,\,(\rho} F^{\,\,\,\,\beta}_{\sigma)}-\frac{(FF)^{(\alpha}_{(\rho} g^{\beta)}_{\sigma)}}{2}\,,
\end{align}
with higher orders being total derivatives $\mathcal U_{(1)}^{\rm s}=i[\partial_q\nabla,\mathcal U_{(0)}^{\rm s}]$, $\mathcal U_{(2)}^{\rm s}=-[\partial_q\nabla,[\partial_q\nabla,\mathcal U_{(0)}^{\rm s}]]/2$. The mixed part of $\mathcal U$ has a decomposition as
\begin{align}\nonumber
\frac{\left[\mathcal U_{(0)}^{\rm mx}\right]^{\alpha\beta}_{\rho\sigma}}{\kappa^2}&=\left(g^{\mu(\alpha}\phi^{,\beta)}-g^{\alpha\beta}\phi^{,\mu}\right)\frac{q_\mu q_{\nu}}{q^2}g^\nu_{(\rho}\phi_{,\sigma)}\\ 
	&-\left(g^{[\lambda(\alpha}F^{\beta)\mu]}-g^{\alpha\beta}F^{\lambda\mu}\right)\frac{q_\mu q^\nu}{q^2}\left(g_{[\lambda(\rho} F_{\sigma)\nu]}-g_{\rho\sigma}F_{\lambda\nu}\right)\,,\label{mxcU0}
\end{align}
for the zeroth order while
\begin{align}\label{mxcU1}
\frac{\left[\mathcal {U}^{\rm mx}_{(1)}\right]^{\alpha\beta}_{\rho\sigma} }{\kappa^2}=
&
\left((\psi_{;\mu})^\dagger\moe^\mu g^{\alpha\beta}-\frac{(\psi^{;(\alpha})^\dagger\moe^{\beta)}}{2}\right)\frac{1}{\sigma\cdot q}(g_{\rho\sigma}\moe^\nu\psi_{;\nu}+\moe_{(\rho}\psi_{;\sigma)})\\ \nonumber
&-\frac{im_\phi^2\phi}{q^2}\left(g^{\alpha\beta} q_{(\rho}\phi_{;\sigma)}+(g^{\mu(\alpha} \phi^{;\beta)}-g^{\alpha\beta}\phi^{;\mu})q_\mu g_{\rho\sigma} \right)\\\nonumber &+i\left(g^{\mu(\alpha}\phi^{,\beta)}-g^{\alpha\beta}\phi^{,\mu}\right)_{;\nu}\left[\partial_q^\nu,\frac{q_\mu q_{(\rho}}{q^2}\right]\phi_{;\sigma)}\\ \nonumber
&-i\left(g^{[\lambda(\alpha}F^{\beta) \mu]}-g^{\alpha\beta}F^{\lambda\mu}\right)_{;\omega}\left[\partial_q^\omega			,\frac{q_\mu q^\nu}{q^2}\right]\left(g_{[\lambda(\rho} F_{\sigma)\nu]}-g_{\rho\sigma}F_{\lambda\nu}\right)\\ \nonumber
&+i\frac{q_\mu q_{\nu}}{q^2}\left(\left(g^{\mu(\alpha}\phi^{,\beta)}-g^{\alpha\beta}\phi^{,\mu}\right)g^\nu_{(\rho}\phi_{,\sigma)}\right)_{;\omega}\partial_q^\omega\\ \nonumber
&-i\frac{q_\mu q^\nu}{q^2}\left(\left(g^{[\lambda(\alpha}F^{\beta) \mu]}-g^{\alpha\beta}F^{\lambda\mu}\right)\left(g_{[\lambda(\rho} F_{\sigma)\nu]}-g_{\rho\sigma}F_{\lambda\nu}\right)\right)_{;\omega}\partial_q^\omega\,,
\end{align}
for first and for second
\begin{align}\label{mxcU2}
\frac{\left[{\mathcal U}_{(2)}^{\rm mx}\right]^{\alpha\beta}_{\rho\sigma}}{\kappa^2}=&\left(g^{\mu(\alpha}\phi^{,\beta)}-g^{\alpha\beta}\phi^{,\mu}\right)\left(\CDv^{(1)}_\mu\frac{ q_{\nu}}{q^2}+ \frac{q_\mu}{q^2}\CDv^{(1)}_{\nu}-\frac{q_\mu}{q^2}\left\{q,\CDv^{(1)}\right\}\frac{q_{\nu}}{q^2}\right)g^\nu_{(\rho}\phi_{,\sigma)}\\ \nonumber
&-\left(g^{[\lambda(\alpha}F^{\beta) \mu]}-g^{\alpha\beta}F^{\lambda\mu}\right)\left(\CDv^{(1)}_\mu\frac{ q^{\nu}}{q^2}+ \frac{q_\mu}{q^2}\CDv_{(1)}^{\nu}-\frac{q_\mu}{q^2}\left\{q,\CDv_{(1)}\right\}\frac{q^{\nu}}{q^2}\right)
\left(g_{[\lambda(\rho} F_{\sigma)\nu]}-g_{\rho\sigma}F_{\lambda\nu}\right)\\  \nonumber
&+\left(g^{[\lambda(\alpha} F^{\beta)\mu]}-g^{\alpha\beta}F^{\lambda\mu}\right)\frac{q_\mu R_{\lambda}^{\,\,\omega}q^\nu}{q^4}\left(g_{[\omega(\rho} F_{\sigma)\nu]}-g_{\rho\sigma}F_{\omega\nu}\right)\\ \nonumber
&-\frac12\left(g^{\mu(\alpha}\phi^{,\beta)}-g^{\alpha\beta}\phi^{,\mu}\right)_{;..}\left[\partial_q^{.2},\frac{q_{\mu}q_{\nu}}{q^2}\right]g^\nu_{(\rho}\phi_{,\sigma)}\\ \nonumber
&-\left(g^{\mu(\alpha}\phi^{,\beta)}-g^{\alpha\beta}\phi^{,\mu}\right)_{;.}\left[\partial_q^.,\frac{q_{\mu}q_{\nu}}{q^2}\right]g^\nu_{(\rho}\phi_{;\sigma)\omega}\partial_q^\omega-\frac{m_\phi^4\phi^2 g^{\alpha\beta}g_{\rho\sigma}}{q^2}\\\nonumber
&+m_\phi^2\left((q^{(\alpha}\phi^{\beta)}-g^{\alpha\beta}(q\phi^;))\frac{1}{q^2}g_{\rho\sigma}\phi_{;\omega}\partial_q^\omega +g^{\alpha\beta}\phi_{;.} \partial_q^. \frac{1}{q^2} q_{(\rho}\phi_{;\sigma)}+(q^{(\alpha}\phi^{\beta)}-g^{\alpha\beta}(q\phi^{;}))\frac{1}{q^4}(q_{(\rho}\phi_{;\sigma)})\right)\\  \nonumber
&+\frac12\left(g^{[\lambda(\alpha}F^{\beta)\mu]}-g^{\alpha\beta}F^{\lambda\mu}\right)_{;..}\left[\partial_q^{.2},\frac{q_{\mu}q^{\nu}}{q^2}\right]
\left(g_{[\lambda(\rho} F_{\sigma)\nu]}-g_{\rho\sigma}F_{\lambda\nu}\right)\\ \nonumber
&+ \left(g^{[\lambda(\alpha}F^{\beta)\mu]}-g^{\alpha\beta}F^{\lambda\mu}\right)_{;.}\left[\partial_q^.,\frac{q_{\mu}q^{\nu}}{q^2}\right]
\left(g_{[\lambda(\rho} F_{\sigma)\nu]}-g_{\rho\sigma}F_{\lambda\nu}\right)_{;\omega}\partial_q^\omega\\\nonumber
&+i\left[\partial_q^\mu,\frac{q^\nu}{q^2}\right]\left((\psi_{;\omega})^\dagger\moe^\omega g^{\alpha\beta}-\frac{(\psi^{;(\alpha})^\dagger\moe^{\beta)}}{2}\right)_{;\mu}\sigma_\nu(g_{\rho\sigma}\moe^\lambda\psi_{;\lambda}+\moe_{(\rho}\psi_{;\sigma)})+({\rm\,total\,derivative})\,.
\end{align}

The last transformation, $e^{i\mathcal V\partial_q}$, together with the definition in eq.~(\ref{DeftcU})  determines $\tcU_{(0)}=\mathcal U_{(0)}-V^2$ where for convenience this combination has been given in eq.~(\ref{ctU0}). In particular since $V$ is itself $q$-independent we allocate $V^2$ to $\tcU^{\rm s}$ and so being explicit
\begin{align}
\tcU_{(0)}^{\rm s}=&\mathcal U_{(0)}^{\rm s} -V^2\,, & \tcU_{(1)}^{\rm s}=&i[\partial_q\nabla,\tcU_{(0)}^{\rm s}]+i[\partial_qV,\tcU_{(0)}^{\rm s}] \,,
\end{align}
and the second order
\begin{align}\label{Ut2eq}
   \tcU_{(2)}^{\rm s}=&i[V\partial_q,i[\partial_q\nabla,\tcU_{(0)}^{\rm s}]] -\frac12[[\partial_q\nabla,[\partial_q\nabla,\tcU_{(0)}^{\rm s}]]-\frac12[V\partial_q,[V\partial_q,\tcU_{(0)}^{\rm s}]]-[[\nabla,V]\partial_q^2,\tcU_{(0)}^{\rm s}]\,,
\end{align}
Meanwhile for the mixed term we have
\begin{align}
\tcU_{(0)}^{\rm mx}&=\tcU^{\rm mx}_{(0)}\,, &\tcU_{(1)}^{\rm mx}&=\mathcal U_{(1)}^{\rm mx}+i[V\partial_q,\mathcal U^{\rm mx}_{(1)}]\,,
\end{align}
and a second order
\begin{align}\label{Ut2eq}
	\tcU_{(2)}^{\rm mx}=&\mathcal U_{(2)}^{\rm mx}+i[V\partial_q,\mathcal U_{(1)}^{\rm mx}] -\frac12[V\partial_q,[V\partial_q,\mathcal U_{(0)}^{\rm mx}]]-[[\nabla,V]\partial_q^2,\tcU_{(0)}^{\rm mx}]\,.
\end{align}
With these transformed operators one is in a position to evaluate the one loop action. 
\section{Evaluation of the operator trace\label{TrLogEv}}
The evaluation has now been cast into the log of the trace of the transformed operator
\begin{align}\label{OpFin}
	e^{i\mathcal V\partial_q}e^{iT}e^{-iqx}\mathcal O e^{iqx}e^{-iT}e^{-i\mathcal V\partial_q}
=-(q+\tilde \CDv)^2+\widetilde{\mathcal U}\,,
\end{align}
where the transformation $e^{iqx}e^{-iT}$ has turned open derivatives into functions of the commutator $[\nabla,\nabla]$ and $e^{i\mathcal V\partial_q}$ has removed a possible linear term in $\nabla$. However just like $\nabla$ did not commute with $\partial_q\,\&\, q$ so does its commutator, $[\nabla,\nabla]$. To illustrate the relevance of this fact let us rearrange the first term in $\CDv$ as
\begin{align}
\CDv_{(1)}=& \frac 14 \left\{\partial_q^\nu,[\nabla_\nu,\nabla_\mu] \right\}+\frac1{12}R^\nu_{\,\,..\mu}\{q_\nu,\partial_q^{.2}\}\\
=&\frac12 \partial_q^\nu [\nabla_\nu,\nabla_\mu]+\frac 14 \left[[\nabla_\nu,\nabla_\mu],\partial_q^\nu \right]+\frac16 R^\nu_{\,\,..\mu}q_\nu\partial_q^{.2}+\frac1{12}R^\nu_{\,\,..\mu}[\partial_q^{.2},q_\nu]\\
=&\frac12 \partial_q^\nu [\nabla_\nu,\nabla_\mu]+\frac13\partial_q^\nu R_{\nu\mu}+\frac16 R^\nu_{\,\,..\mu}q_\nu\partial_q^{.2}\,.
\label{CDvAllRH}
\end{align}
In this way the commutator acts solely  on whatever lies to the right of $\CDv_{(1)}$. The case for $\tCDv_{(1)}$ is not qualitatively different but for completeness it is
\begin{align}
&\tCDv_{(1)}=\partial_q^\nu\left(\frac12([\nabla_\nu,\nabla_\mu]+\nabla_{[\nu} V_{\mu]}+V_{[\nu}V_{\mu]})-\frac12 \nabla_{(\nu} V_{\mu)}+\frac13 R_{\nu\mu} \right)+\frac{R^\rho_{\,\,\,..\mu}}{6}q_\rho\partial_q^{.2}\,.
\end{align}
When the commutator is acting on the field we are integrating over, i.e. $[\nabla_\nu,\nabla_\mu]$ is to its rightmost in the operator of eq.~(\ref{OpFin}), one has depending on the spin of the field, 
\begin{align}
[\nabla_\alpha,\nabla_\beta] \phi &=0 
&[\nabla _{\alpha},\nabla_{\beta}]\psi&=\frac{\sigma^{[a}\bar\sigma^{b]}}8e_{a,\rho}e^\lambda_{b} R^{\rho}_{\,\,\lambda\alpha\beta}\psi\\
[\nabla_\alpha,\nabla_\beta] A^\mu &=R^\mu_{\,\,\rho \alpha\beta}A^\rho & [\nabla_\alpha,\nabla_\beta] T^{\mu\nu}&= R^{\mu}_{\,\,\rho\alpha\beta}T^{\rho\nu}+R^{\nu}_{\,\,\rho\alpha\beta}T^{\mu\rho}\,,
\end{align}
so it is useful to define
\begin{align}
[\nabla_\alpha,\nabla_\beta]({\rm Field\,}\Phi)\equiv\mathscr{R}_{\alpha\beta}({\rm Field\,}\Phi)\,.
\end{align}
In a way analogous to creation and annihilation operator rearrangement one can put in the form of eq.~(\ref{CDvAllRH}) all terms in the expansion, i.e. the commutator $[\nabla,\nabla]$ to its rightmost position and all $\partial_q$ to the right of $q$'s, e.g. the first order in $\{q,\mathcal K\}$ in this form
\begin{align}
\mathcal O\supset \left\{q,\mathcal K_{(1)}\right\}=&-\frac R 6+q^\mu\partial_q^\nu\left(\mathscr R_{\nu\mu}+\frac 13 R_{\nu\mu} \right)+\frac 13 R^{\star\,\,\,\,\star}_{\,\,\,..}q_\star^2 \partial_q^{.2}\,,
\end{align}
where as for $\partial_q$ the notation $R^{\star\,\,\,\,\star}_{\,\,\,..}q_\star^2=R^{\mu\,\,\,\,\nu}_{\,\,\,..}q_\mu q_\nu$ whereas for the tilded case
\begin{align}
\mathcal O\supset \left\{q,\widetilde{\CDv}_{(1)}\right\}=&-\frac R 6-\nabla V+q^\mu\partial_q^\nu\left(\tilde{\mathscr R}_{\nu\mu}+\frac 13 R_{\nu\mu}-\nabla_{(\nu} V_{\mu)} \right)+\frac 13 R^{\star\,\,\,\,\star}_{\,\,\,..}q_\star^2 \partial_q^{.2}\,.
\label{AllLK}
\end{align}
where we have defined
\begin{align}
	\tilde{\mathscr {R}}_{\mu\nu}=\mathscr R_{\mu\nu}+\nabla_{[\mu} V_{\nu]}+ V_{[\mu}V_{\nu]}\,,
\end{align}
the fact that this structure arranges as $[\nabla+V,\nabla+V]$ suggests a combined transformation in place of $e^{iT}e^{i\partial_q \mathcal V}$ might simplify the algebra. Nevertheless here such option is not pursued since in contrast to the universal $\nabla$, the action of $V$ might be confined to a single operator.

In the form of eq.~(\ref{AllLK}) the hermiticity is not an obvious property yet it is more adequate for computations since all commutators are `evaluated' as opposed to $\partial_q$, for whom it is still left to specify what is acts on. 
For this purpose let us rewrite the one loop correction introducing $m^2$, (not to be confused with the scalar mass $m_\phi^2$) the one loop action of eq.~(\ref{Origin}):
\begin{align} \nonumber
\frac{i}{2}\mbox{tr}\log(\mathcal O+m^2)=&\frac i2
\int \frac{d^dxd^d q}{(2\pi)^d}\int dm^2\mbox{tr}[(\mathcal O+m^2)^{-1}] \\ 
=& \frac{i}{2}\int \frac{d^dxd^d q}{(2\pi)^d}\int dm^2\mbox{tr}[(-q^2+m^2-\{\tCDv,q\}-\tCDv^2+ \mathcal U)^{-1}]\\ \nonumber
=&-\frac{i}{2}\int \frac{d^dxd^d q}{(2\pi)^d}\int dm^2\sum\mbox{tr}\left(\left[\frac{1}{q^2-m^2} (\tcU-\{q,\tCDv\}-\tCDv^2)\right]^n \frac{1}{q^2-m^2}\right)
\end{align}
where $m^2$ will, at the end of the calculation be taken to $0$ but in general it is useful to keep it as an IR regulator as indeed not all terms converge for $m^2\to0$ and the order of integration shall be kept as above.
Once all terms in $\tilde\CDv, \tcU$ are in the form of eq.~(\ref{AllLK}) only $\partial_q$ is left to act on propagators and other terms in the expansion to its right. After allowing all $\partial_q$ to make their way to the right the result will be  momenta $q$ contracted with Lorentz tensors made out of the background fields. The momentum dependence in $q$ after loop integration will yield tensors built out of the metric (recall $q$ is a covariant object $q_\mu$, $q^2=q_\mu q_\nu g^{\mu\nu}$).

With our expansion of $\tilde\CDv, \tcU$ in its dimensions in loop momenta we can organize the effective action; the first order is ${\color{blue}\mathcal O(q^{d-2})}$;
\begin{align}
&\mbox{tr}\log(\mathcal O+m^2)\\\nonumber
=&\int \frac{d^dxd^d q dm^2}{(2\pi)^d} \left(\frac{1}{q^2-m^2}\left(\widetilde{\mathcal U}_{(0)}-\{q,\widetilde\CDv_{(1)}\}\right)\frac{1}{q^2-m^2}
\right)+\mathcal{O}(q^{d-4})
\end{align}
Taking for demonstration a scalar field
and with the result in eq.~(\ref{AllLK}) 
\begin{align}
	&\int \frac{d^dxd^dq}{(2\pi)^d} \int dm^2\frac{1}{q^2-m^2}\left(U_{(0)}-\{q,\CDv_{(1)}\}\right)\frac{1}{q^2-m^2}\\ \nonumber
	=&\int \frac{d^dxd^dq}{(2\pi)^d} \int dm^2\frac{1}{q^2-m^2}\left(\frac 1 6 R  -\left(\mathscr R_{.*}+\frac 13 R_{.*} \right)q^*\partial_q-\frac 13 R^{\star\,\,\,\star}_{\,\,\,..} q_\star^2\partial_q^{.2}\right)\frac{1}{q^2-m^2}\\ \nonumber=&\int d^dx\frac R 6\int \frac{d^dq}{(2\pi)^d(q^2-m^2)}
\end{align}
which for dimensional regularization is non vanishing (when $m^2\to0$) only for $d= 2$ and contributes for scalars the well-known $N_S/(24\pi)$ to Weyl's anomaly (the `$-26/24\pi$' contribution for the bosonic string we cannot reproduce since Weyl scaling was not taken as local symmetry).
The focus of this paper is however $d=4$ and the UV divergences contained in the next non-vanishing order ${\color{blue}\mathcal O(q^{d-4})}$:
\begin{align} \label{allofthem}
&	\mbox{tr}\log(\mathcal O+m^2)=\mathcal O(q^{d-2})\\ \nonumber 
+&\boxed{\int \frac{d^dxd^dq}{(2\pi)^d} \int dm^2\left( \Prp\left(\widetilde{\mathcal U}_{(2)}-\{q,\widetilde\CDv_{(3)}\}-\widetilde\CDv_{(1)}^2\right)+\left(\Prp\left(\widetilde{\mathcal U}_{(0)}-\{q,\widetilde\CDv_{(1)}\}\right)\right)^2\right)
\Prp}+\mathcal{O}(q^{d-6})
\end{align}
where for brevity we introduced $\Delta=(q^2-m^2)^{-1}$ and this is the integral at the core of our computation. This expression, safe for the term $\tilde\CDv_{(3)}$, resembles the static flat background case~\cite{Henning:2014wua} taking loosely speaking $\mathcal K$ as our (field strength)$\times\partial_q $.

Given the main novel result of this work, i.e. the covariant derivative in eq.~(\ref{FistCDE}), eq.~(\ref{allofthem}) can be evaluated
in a straight-forward way as done for the $\mathcal O(q^{d-2})$ term as sketched above and in particular the UV terms can be computed with the regularization of choice.
The amount of algebra now nonetheless makes it more digestible to split the computation into sections and introduce some minimal notation. Here dimensional regularization will be employed and the following definition for an integral and propagator 
\begin{align}\label{DefmqInt}
&\mqInt\equiv \underset{d\to4}{\rm lim}\frac{8\pi^2(4-d)}{i\sqrt{|g|}}\int \frac{d^dq dm^2}{(2\pi)^d}\,,
&&\Delta\equiv\frac{1}{q^2-m^2}\,,
\end{align}
casts the UV contributions subject of this work as
\begin{align}
\mathscr L_{\rm UV}&=\frac{1}{(4\pi)^2(4-d)}\mqInt \sum\mbox{tr}\left(\left(\Delta (\tcU-\{q,\tCDv\}-\tCDv^2)\right)^n \Delta\right)\\
&\equiv \frac{1}{(4\pi)^2(4-d)} \mqInt\left(\mathcal I_{\rm s}+\mathcal I_{\rm mx}\right)\,,\label{FinAct}
\end{align}
where
\begin{align} \label{SingleSpecies}
\mathcal I_{\rm s}=&\left( \Prp\left(\widetilde{\mathcal U}_{(2)}^{\rm s}-\{q,\widetilde\CDv_{(3)}\}-\widetilde\CDv_{(1)}^2\right)+\left(\Prp\left(\widetilde{\mathcal U}_{(0)}^{\rm s}-\{q,\widetilde\CDv_{(1)}\}\right)\right)^2\right)
\Prp\,,\\ \label{Mixed}
\mathcal I_{\rm mx}=& \left(\Prp \tcU_{(2)}^{\rm mx}+\left(\Prp \tcU^{\rm mx}_{(0)}\right)^2+\left\{\Prp \tcU^{\rm mx}_{(0)}\,,\Prp\left( \widetilde{\mathcal U}_{(0)}^{\rm s}-\{q,\widetilde\CDv_{(1)}\}\right)\right\}\right)\Prp\,,
\end{align}
encode the contributions from single-spin species running in the loop and mixed contributions respectively.
The following sections are concerned with the part of the effective action computation for each of these to cases: single species loops~\ref{sec:GRren} $\mathcal I_{\rm  s}$,  and mixed-species loops~\ref{sec:MatRen}, $\mathcal I_{\rm  mx}$.

\subsection{Single species loops\label{sec:GRren}}
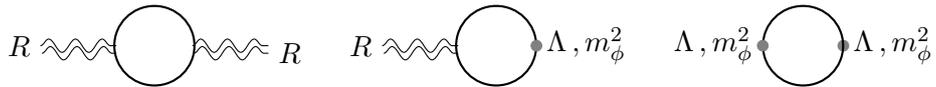
\begin{figure}[h]\centering
\begin{tikzpicture}
	\draw [style={decorate, decoration={snake}}] (0,0) node [anchor=east] {$R$} -- (1,0);
		\draw [style={decorate, decoration={snake}}] (.02,-.1) -- (1,-.1);
			\draw [style={decorate, decoration={snake}}] (2,0) -- (3,0);
		\draw [style={decorate, decoration={snake}}] (2.02,-.1) -- (3,-.1) node [anchor=west] {$R$};
		\draw[thick] (1.5,0) circle (15pt);
\end{tikzpicture}\quad
		\begin{tikzpicture}
		\draw [style={decorate, decoration={snake}}] (-1,0) node [anchor=east] {$R$} -- (0,0);
		\draw [style={decorate, decoration={snake}}] (-.98,-.1) -- (0,-.1);
		\draw[thick] (.5,0) circle (15pt);
		\filldraw[gray] (1.03,-0) circle (2pt) node [anchor=west, black] {$\Lambda\,,m_\phi^2$};
		\end{tikzpicture}\quad
			\begin{tikzpicture}
\filldraw[gray] (2.03,-0) circle (2pt) node [anchor=west, black] {$\Lambda\,,m_\phi^2$};
	\draw[thick] (1.5,0) circle (15pt);
			\filldraw[gray] (.97,0) circle (2pt) node [anchor=east, black] {$\Lambda\,,m_\phi^2$};
	\end{tikzpicture}
	\caption{Schematic of the UV divergent curvature terms at one loop\label{fig1}}
\end{figure}

The integration of a given spin field results in the UV divergent terms of  eqs.~(\ref{FinAct}) with
\begin{align}
\mathcal I_{\rm s}=\Prp\left(\tcU_{(2)}^{\rm s}-\{q,\tCDv_{(3)}\}-\tCDv_{(1)}^2\right)\Prp+\left(\Prp\left(\tcU_{(0)}^{\rm s}-\{q,\tCDv_{(1)}\}\right)\right)^2\Prp\,, \label{MstSP}
\end{align}
this subsection carries out the loop integrals and yields the 1-loop corrections.

Let us start with
\begin{align}
	\mqInt \Prp{\rm tr}( \tcU_{(2)}^{\rm s})\Prp\,,
\end{align}
here total derivatives are neglected and hence the $-[\nabla,[\nabla, \tcU^{\rm s}_{(0)}]]\partial_q^2/2$ (with ${\tcU}^{\rm s}_{(0)}=U^{\rm s}-V^2$) piece in $\mathcal U_{(2)}^{s}$ as per eq.~(\ref{Ut2eq}) can be ignored, whereas for the remainder of $\tcU_{2}$ 
\begin{align}\nonumber
	{\tcU}^{\rm s}_{(2)}+[\nabla,[\nabla, \tcU^{\rm s}_{(0)}]]\frac{\partial_q^2}{2}&=
	 i[V\partial_q, i[\nabla,{\tcU}^{\rm s}_{(0)}]\partial_q]-\frac12[V\partial_q,[V\partial_q,{\tcU}^{\rm s}_{(0)}]]-[[\nabla,V]\partial_q^2,{\tcU}^{\rm s}_{(0)}]\\
	&=
	 i[V, i[\nabla,{\tcU}^{\rm s}_{(0)}]]\partial_q^2-\frac12[V,[V,{\tcU}^{\rm s}_{(0)}]]\partial_q^2-[[\nabla,V],{\tcU}^{\rm s}_{(0)}]\partial_q^2\Prp
\end{align}
where we used that $[\partial_q,\tcU^{\rm s}_{(0)}]=[\partial_q,V]=0$ in the second line. This form makes clear that these are commutators of matrices which yield zero when traced over. 
One has that for the mixed pieces $[\partial_q,\mathcal U^{\rm mx}_{(i)}]\neq 0$ and this terms do contribute, as made explicit in sec.~\ref{sec:MatRen}.
On the other hand the results of tilding $\CDv_{(3)}$ are terms which vanish when tracing over them or of the form of
\begin{align}
\mqInt \Prp\{ q, (\tCDv_{(3)}-\CDv_{(3)})\}\Prp\supset \mqInt \Prp \{q,i[V\partial_q,\CDv_{(2)}]\}\Prp
\end{align}
where
\begin{align}
[V\partial_q,\CDv^{(2)}_\mu]=-\frac 16 \{\left[V\partial_q,\left[\partial_q\nabla,[\nabla_\nu,\nabla_\mu]\right]\right],\partial_q^\nu\}-\frac1{24}R^{\nu}_{\,\,\,..\mu;.}\{\partial_q^{.3},V_\rho\left[\partial_q^\rho,q_\nu\right]\}
\end{align}
which, regardless of the matrix structure contained, involve the vanishing integral
\begin{align}
	\mqInt \Prp \{q,\partial_q^3\}\Prp=0\,,
\end{align}
and so one can drop the tilde and consider $\mathcal K_{(3)}$ only.
Given these cancellations and total derivative terms the part relevant of eq.~(\ref{MstSP})  is:
\begin{align}\nonumber
&\Prp\left(-\{q,\CDv_{(3)}\}-\tCDv_{(1)}^2\right)\Prp+\left(\Prp\left(\tcU_{(0)}^{\rm s}-\{q,\tCDv_{(1)}\}\right)\right)^2\Prp\\	 \label{CDv3term}
=
&-\Prp\left\{q^\mu,\frac{1}{48}\{R^{\nu}_{\,\,..\mu}\partial_q^{.3},[\nabla_.,\nabla_\nu]\}+\frac{7}{720}\{(R^\nu_{\,\,..\rho}R^{\rho}_{\,\,..\mu}\partial_q^{.4},q_\nu\}\right\}\Prp\\  \nonumber
&-\Prp\left(
\partial_q^\nu\left(\frac12([\nabla_\nu,\nabla_\mu]+\nabla_{[\nu} V_{\mu]}+V_{[\nu}V_{\mu]})-\frac12 \nabla_{(\nu} V_{\mu)}+\frac13 R_{\nu\mu} \right)+\frac{R^\nu_{\,\,..\mu}}{6}q_\nu\partial_q^{.2}
\right)^2\Prp\\ \nonumber
&+\left(\Prp\left(\tcU_{(0)}^{\rm s}+\frac 1 6 R  +\nabla V-\left(\widetilde{\mathscr R}_{\nu\mu}+\frac 13 R_{\nu\mu}-\nabla_{(\nu} V_{\mu)} \right)q^\mu\partial_q^\nu-\frac  {R^{*\,\,\,*}_{\,\,..}}3 q^2_*\partial_q^{.2}\right)\right)^2\Prp\,.
\end{align}
Here the detailed loop integral computation is not made explicit for all terms, rather it is carried out for the first term of eq.~(\ref{CDv3term}) since this is the novel term that differs with the flat metric case.
First, via the relation
\begin{align}
\{A,\{B,C\}\}=\{\{A,B\},C\}+[B,[C,A]]=2\{A,B\}C+[C,\{A,B\}]+[B,[C,A]]
\end{align}
one has, making all $q$ dependence explicit,
\begin{align}
\frac7{720}R^\mu_{\,\,..\rho}R^{\rho\,\,\,\nu}_{\,\,..}\int\frac{d^dq dm^2}{(2\pi)^d} &\frac{1}{q^2-m^2}\left(4 q_\mu q_\nu \partial_q^{.4}+2g_{(\nu.}q_{\mu)}\partial_q^{.3}+g^{.}_\nu  g^._\mu\partial_q^{.2}\right)\frac{1}{q^2-m^2}\\ \nonumber
= \frac7{720}R^\mu_{\,\,..\rho}R^{\rho\,\,\,\nu}_{\,\,..}\int\frac{d^dq dm^2}{(2\pi)^d}\Bigg(&-2\frac{g_{\nu}^.g_{\mu}^.g^{..}_{(\times \rm \color{purple} 12)}}{(q^2-m^2)^3}\\ \nonumber
 &+\frac{8}{(q^2-m
^2)^4}\left(g_{\nu}^.g_{\mu}^.q^{.2}_{(\times \rm \color{purple} 12)}+2g_{(\nu .}q_{\mu)}q^.g^{..}_{(\times \rm \color{purple} 12)}+4q_\nu q_\mu g^{..}g^{..}_{(\times \rm \color{purple} 3)}\right)\\ \nonumber
&-\frac{96}{(q^2-m^2)^5}\left(g_{(\nu}^.q_{\mu)}q^{.3}_{(\times \rm \color{purple} 4)}+2q_\nu q_\mu q^{.2}g^{..}_{(\times \rm \color{purple} 12)}\right)+1536\frac{q_\nu q_\mu q^{.4}}{(q^2-m^2)^6}\Bigg)\\
=\frac7{720}\frac13\left(R_{..}^2+\frac32R_{....}^2\right)&\int\left(\frac{d^dq}{(2\pi)^dq^4}+\mathcal O\left(\frac{m^2}{q^6}\right)\right)
\end{align}
where $R_{....}^2=R_{\alpha\beta\gamma\delta}R^{\alpha\beta\gamma\delta}$, the purple subscript indicates the multiplicity in terms from symmetrizing in `$.$' indices and we used $R_{\alpha\beta\gamma\delta}R^{\alpha\gamma\beta\delta}=R_{....}^2/2$. Even if somewhat involved the contrast with conventional Feynman-diagram techniques makes this integral, the basic element of the computation, a relatively simple exercise whereas no knowledge of the heat-kernel method or De-Witt coefficients was required.

The other term in $\CDv_{(3)}$ adds up with the above to yield:
\begin{align}
\int\frac{d^dq dm^2}{(2\pi)^d}\Prp\left\{q,\CDv_{(3)} \right\}\Prp=\left(\frac7{720}-\frac1{48}\right)\frac13\left(R_{..}^2+\frac32R_{....}^2\right)&\int\frac{d^dq}{(2\pi)^dq^4}+\mathcal O\left(\frac{m^2}{q^2}\right)
\end{align}

The loop integration for the left-over terms in~(\ref{CDv3term}) follows the above lines and results in, with the abbreviated notation of~(\ref{DefmqInt}), one of the main results here derived
\begin{align}\label{KurRes}
\mqInt\, \mathcal I_{\rm s}=&\mqInt\left(	\Prp\left(-\{q,\tCDv_{(3)}\}-\tCDv_{(1)}^2\right)\Prp+\left(\Prp\left(\tcU_{(0)}^{\rm s}-\{q,\tCDv_{(1)}\}\right)\right)^2
	\Prp\right) \\ \nonumber
&=	\left(\frac{R_{....}^2}{180}-\frac{R_{..}^2}{180}\right)\mbox{tr}(\mathbb I)+\frac{1}{12}\mbox{tr}\left(\tilde{\mathscr R}_{\mu\nu}\tilde{\mathscr R}^{\mu\nu}\right)+\frac{1}{2}\mbox{tr}\left(\tcU_{(0)}^{\rm s}+\frac R6\right)^2+{(\rm  total\,\, der.)} 
\end{align}
This 1 loop result has long been available in the literature, see~\cite{Fradkin_1977,Barvinsky:1985an,Buchbinder:1992rb}, yet the emphasis here is the new computational technique.
In this regard the universal formulae for the flat case taking $[F_{\mu\nu}]^a_{\,\,b}\to R^{\alpha}_{\,\,\beta \mu\nu}$ reproduces all terms except the first one which `counts' the degrees of freedom, is connected to the $a$ theorem and has been explicitly computed here. If one splits the contribution by the dimension of the operators, for the action of eq.~(\ref{OrAct}) and according to eq.~(\ref{ctU0}) the sum runs from a CC term to dimension twelve (see~\cite{Ruhdorfer:2019qmk} for a study of the operator basis) which here we organize as
\begin{align}
\mathcal I_{\rm s}= \sum_{n=0}^6\kappa^{2n-4}\mathcal I_{\rm s}^{2n}(\alpha_{m_\phi},\alpha_\Lambda,R,\phi,\psi,F)
\end{align}  
where the action taken as a function of only one dimensionfull parameter $\kappa^{-1}= M_{\rm pl}/\sqrt{8\pi}$ and ratios $\alpha_{m_\phi}\equiv m_\phi^2\kappa^2$, $\alpha_\Lambda\equiv\Lambda\kappa^2$. A set of diagrams, which although incomplete represents all the possible external fields is given in figs~\ref{fig1}-\ref{fig3}.

Let us look at the curvature square $R^2$ terms explicitly caring for the ghosts contributions as well in the structure of eq.~(\ref{KurRes}):
\begin{align} \label{R2UV}
	{\rm Field} && &{\rm tr}(\mathbb{I})(R_{....}^2-R_{..}^2)/180&&{\rm tr}(\mathscr R^2)/12&&{\rm tr}(\tcU_{(0)}^2)/2\\[1pt] \hline \nonumber &&&&&&&\\[-10pt] \nonumber
{\rm Ghost}(c^\mu) & &(-2)\Bigg[&4\left(\frac{R_{....}^2}{180}-\frac{R_{..}^2}{180}\right) &+&\frac{1}{12}(-R_{....}^2)&+&\frac12\left(R_{..}^2+\frac49R^2\right)\Bigg]\\ \nonumber
{\rm Metric} & & &10\left(\frac{R_{....}^2}{180}-\frac{R_{..}^2}{180}\right)&+&\frac{1}{12}(-6R_{....}^2)&+&\frac12\left(3R_{....}^2-4R_{..}^2+\frac{22}{36}R^2\right)\\ \nonumber
{\rm Scalar} &  &&\left(\frac{R_{....}^2}{180}-\frac{R_{..}^2}{180}\right)&&&+&\frac12\left(\frac{R}{6}\right)^2\\\nonumber
{\rm CFT~Scalar} && &\left(\frac{R_{....}^2}{180}-\frac{R_{..}^2}{180}\right) &&&& \\\nonumber {\rm Weyl\,Fermion} & &(-1)\Bigg[&2\left(\frac{R_{....}^2}{180}-\frac{R_{..}^2}{180}\right)&+&\frac{1}{12}\frac{(-1)}4 R_{....}^2
&+&\frac12\frac {R^2}{72}\,\,\Bigg]\\ \nonumber 
{\rm Vector\,boson}& && 4\left(\frac{R_{....}^2}{180}-\frac{R_{..}^2}{180}\right)&+&\frac1{12}(-R_{....}^2)&+&\frac12\left(R_{..}^2-\frac29R^2\right)\\ \nonumber
{\rm  Ghost} (c)&&(-2)&\Bigg[\left(\frac{R_{....}^2}{180}-\frac{R_{..}^2}{180}\right)&&&+&\frac12\left(\frac{R}{6}\right)^2\,\,\Bigg]
\end{align}
If there are $N_\phi$ scalars, $N_\psi$ fermions and $N_A$ (spin 1) gauge bosons the contribution reads
\begin{align}\nonumber
	R_{....}^2\left(\frac{N_\phi-13N_A}{180}+\frac{7N_\psi}{720} \right)+R_{..}^2\left(\frac{2N_\psi-N_\phi+88N_A}{180}\right)+R^2\left(\frac{2N_\phi-N_\psi-20N_A}{144}\right)
\end{align}
and so for the SM input $N_i=\{4,45,12\}$. One can also project onto the basis of Euler number density ($\tilde R_{....}^2=R_{....}^2-4R_{..}^2+R^2$) and Weyl tensor ($C_{....}^2=R_{....}^2-2R_{..}^2+R^2/3$) and a total derivative ($\nabla J=R_{....}^2+R_{..}^2+3R^2$) with the transformation
\begin{align}\left(\begin{array}{c}c_{\tilde R}\\
	c_{C}\\c_{\nabla J}\end{array}\right)=
\frac1{22}\left(\begin{array}{ccc}
39&6&-15\\
-19&-8&9\\
2&2&6
\end{array}\right)\left(\begin{array}{c} c_{R_{....}}\\c_{R_{..}}\\c_R\end{array}\right)
\end{align}
for the coefficients of each operator to check that the trace anomaly is reproduced as in e.g.~\cite{Duff:1993wm}. 
\begin{figure}\centering
	\begin{tikzpicture}
	\draw  (0.22,0.33) -- (1,0);
	\draw  (0.22,-0.33) node [anchor=south east] {$T$} -- (1,0);
	\draw  (2,0) -- (2.78,0.33) node [anchor=north west]{$T$};
	\draw  (2,0) -- (2.78,-0.33);
	\draw[style={decorate, decoration={snake}}] (1.530,0) circle (15pt);
	\draw[style={decorate, decoration={snake}}] (1.470,0) circle (15pt);
	\draw (0,-.8) node {$.$};
	\end{tikzpicture}\qquad
	\begin{tikzpicture}
	\draw  (0.5,.75) node [anchor=south east] {$T$} -- (1.133,1/2);
	\draw  (1,1.15) -- (1.133,1/2);
	\draw  (2,1.15)   -- (1.866,1/2);
	\draw  (2.6,.75) node [anchor=south west] {$T$} -- (1.866,1/2);
	\draw  (1.533,-.41) -- (1.2,-1);
	\draw  (1.5,-.41) -- (1.8,-1) node [anchor=south west] {$T$};
	\draw[style={decorate, decoration={snake}}] (1.530,0) circle (15pt);
	\draw[style={decorate, decoration={snake}}] (1.470,0) circle (15pt);
	\end{tikzpicture}\qquad 
	\begin{tikzpicture}
	\draw  (0.5,.75) node [anchor= north east] {$T$}-- (1.1465,.3535);
	\draw  (1,1.1) -- (1.1465,.3535);
	\draw  (2,1.1)  -- (1.8535,.3535);
	\draw  (2.5,.75) node [anchor=north west] {$T$} -- (1.8535,.3535);
	\draw  (0.5,-.75) node [anchor=south east] {$T$}-- (1.1465,-.3535);
	\draw  (1,-1.1) -- (1.1465,-.3535);
	\draw  (2.1,-1.1)  -- (1.8535,-.3535);
	\draw  (2.5,-.75) node [anchor=south west] {$T$}-- (1.8535,-.3535);
	\draw[style={decorate, decoration={snake}}] (1.530,-.1) circle (15pt);
	\draw[style={decorate, decoration={snake}}] (1.470,-.1) circle (15pt);
	\end{tikzpicture}
	\caption{Schematic of UV divergent matter terms at one loop where T stands for the stress energy tensor\label{fig2} so schematically $T\sim \phi^2+\psi^2+F^2$.}
\end{figure}
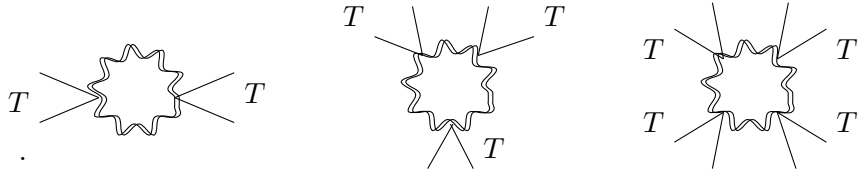
\begin{figure}[h]\centering
	\begin{tikzpicture}
	\draw  (0.22,0.4) node [anchor=north east] {$T$} -- (1,0);
	\draw  (0.22,-0.4) -- (1,0);
	\draw [style={decorate, decoration={snake}}] (2,0) -- (3,0);
	\draw [style={decorate, decoration={snake}}] (2.02,-.075) -- (3,-.075) node [anchor=west] {$R$};
	\draw[style={decorate, decoration={snake}}] (1.530,0) circle (15pt);
	\draw[style={decorate, decoration={snake}}] (1.470,0) circle (15pt);
	\draw (0,-1) node {\,};
	\end{tikzpicture}\qquad
	\begin{tikzpicture}
	\draw  (0.22,0.4) node [anchor=north east] {$T$} -- (1,0);
	\draw  (0.22,-0.4)  -- (1,0);
	\draw[style={decorate, decoration={snake}}] (1.530,0) circle (15pt);
	\draw[style={decorate, decoration={snake}}] (1.470,0) circle (15pt);
	\filldraw[gray] (2,0) circle (2pt) node [anchor=west, black]{$\,\,\,\Lambda$};
	\draw (0,-1) node {\,};
	\end{tikzpicture}\qquad
	\begin{tikzpicture}
	\draw  (0.5,.75) node [anchor=south east] {$T$} -- (1.133,1/2);
	\draw  (1,1.15) -- (1.133,1/2);
	\draw  (2,1.15)   -- (1.866,1/2);
	\draw  (2.6,.75) node [anchor=south west] {$T$} -- (1.866,1/2);
	\draw [style={decorate, decoration={snake}}] (1.47,-.40) -- (1.45,-1);
	\draw [style={decorate, decoration={snake}}] (1.53,-.42) -- (1.55,-1.02) node [anchor=west] {$R$};
	\draw[style={decorate, decoration={snake}}] (1.530,0) circle (15pt);
	\draw[style={decorate, decoration={snake}}] (1.470,0) circle (15pt);
	\end{tikzpicture}
	\caption{Schematic of the UV divergent terms at one loop\label{fig3}}
\end{figure}
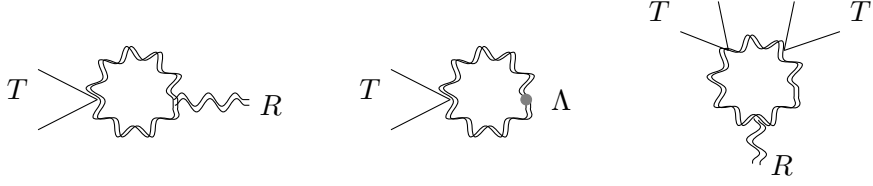
The remaining terms are contained in $\mathscr R^2$ or $(\tcU+R/6)^2$ and are straightforward to obtain. Here we do not reproduce them all but give for scope the lowest dimensional operators generated
\begin{align}
\mqInt \,(\mathcal I_{\rm s}^{0}+\mathcal I_{\rm s}^{2})=\frac 1 2 m_\phi^4{\rm tr}(\mathbb{I}_\phi)+5\Lambda^2+\left(\frac{m_\phi^2}{6}{\rm tr}(\mathbb{I}_\phi)+\frac43\Lambda\right) R+8\Lambda m_{\phi}^2\kappa^2\phi^2\label{RpLUV}
\end{align}
where this contribution together with those in eq.~(\ref{R2UV}) encapsulates all spin $\leq 1$ contributions and on the other end the highest dimensional term generated is
\begin{align}
\mqInt \,\mathcal I_{\rm s}^{12}=\frac{45\kappa^8}{2048}\left(\psi^\dagger \sigma \psi\right)^4\,,
\end{align}
which produces an 8-point amplitude that grows with energy E as $\kappa^8E^4$.
\subsection{Mixed contributions in the loop \label{sec:MatRen}}
\begin{figure}[h]\centering
	\begin{tikzpicture}
	\draw  (0,0) node [anchor=south west] {$\sqrt{T}$} -- (1,0);
	\draw  (2,0) -- (3,0) node [anchor=north east]{$\sqrt{T}$};
	\draw[thick] (2,0) arc (0:180:15pt);
	\draw[style={decorate, decoration={snake}}] (2.025,0) arc (0:-180:15pt);
	\draw[style={decorate, decoration={snake}}] (1.975,0) arc (0:-180:15pt);
	\draw (0,-.8) node {$.$};
	\end{tikzpicture}\quad
	\begin{tikzpicture}
\draw  (-1.7,1) node [anchor=north east] {$\sqrt{T}$} -- (-0.866,1/2);
\draw  (0.,1/2) -- (.86,1);
\draw [style={decorate, decoration={snake}}] (-0.44,-.37) -- (-.44,-1);
\draw [style={decorate, decoration={snake}}] (-0.39,-.40) -- (-.39,-1) node [anchor=west] {$R$};
\draw[style={decorate, decoration={snake}}] (-0.83,1/2) arc (150:390:15pt);
\draw[style={decorate, decoration={snake}}] (-0.89,1/2) arc (150:390:15pt);
\draw[style=thick] (0.05,1/2) arc (30:150:15pt);
\end{tikzpicture}\quad
\begin{tikzpicture}
\draw[style={decorate, decoration={snake}}] (.77,0) arc (0:45:.75);
\draw[style={decorate, decoration={snake}}] (.77,0) arc (0:-45:.75);
\draw[style={decorate, decoration={snake}}] (.72,0) arc (0:45:.75);
\draw[style={decorate, decoration={snake}}] (.72,0) arc (0:-45:.75);
\draw[style=thick] (0.53,.53) arc (45:135:.75);
\draw[style={decorate, decoration={snake}}] (-.77,0) arc (180:225:.75);
\draw[style={decorate, decoration={snake}}] (-.77,0) arc (180:135:.75);
\draw[style={decorate, decoration={snake}}] (-.72,0) arc (180:225:.75);
\draw[style={decorate, decoration={snake}}] (-.72,0) arc (180:135:.75);
\draw[style=thick] (-0.53,-.53) arc (225:315:.75);
\draw  (-1,1) node [anchor= north east] {$\sqrt{T}$}-- (-0.53,0.53);
\draw  (-1,-1) node [anchor= south east] {$\sqrt{T}$}-- (-0.53,-0.53);
\draw  (1,1) node [anchor= north west] {$\sqrt{T}$}-- (.53,.53);
\draw  (1,-1) node [anchor= south west] {$\sqrt{T}$}-- (.53,-.53);
\end{tikzpicture}
	\caption{Non-exhaustive set of diagrams for mixed contributions\label{FigMx}}
\end{figure}
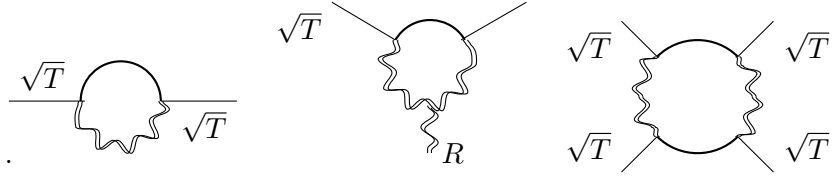
Diagrams with internal particles of different spin contribute terms like those in fig.~\ref{FigMx} and the UV divergences that they give rise to in the effective action read
\begin{align}
\mqInt\mathcal I_{\rm mx}=\mqInt \left(\Prp \tcU^{\rm mx}_{(2)}+\left\{\Prp\left(\tcU_{(0)}^{\rm s}-\left\{q,\tCDv_{(1)}\right\}\right)\,,\Prp \tcU^{\rm mx}_{(0)}\right\}+\left(\Prp \tcU_{(0)}^{\rm mx}\right)^2 \right)\Prp\,.
\end{align}
Let us first address the the $\tcU_{(2)}$ term which is given in terms of $\mathcal U$ in eqs.~(\ref{mxcU0}-\ref{mxcU2})
\begin{align}
\tcU_{(2)}^{\rm mx}=\mathcal U_{(2)}^{\rm mx}+ i[V\partial_q, \mathcal U^{\rm mx}_{(1)}]]-\frac12[V\partial_q,[V\partial_q,\mathcal  U^{\rm mx}_{(0)}]]-[[\nabla,V]\partial_q^2,\mathcal  U^{\rm mx}_{(0)}]\,.
\end{align}
Tracing over these operators one can simplify to
\begin{align}\label{mxtcU2}
	{\rm tr}(\tcU_{(2)}^{\rm mx}-\mathcal U_{(2)}^{\rm mx})={\rm tr}\left( iV[\partial_q, \mathcal U^{\rm mx}_{(1)}]]-\frac12[V\partial_q,[V\partial_q,\mathcal  U^{\rm mx}_{(0)}]]-[\nabla,V][\partial_q^2,\mathcal  U^{\rm mx}_{(0)}]\right)\,,
\end{align}
since for algebraic commutators like $[V,\tcU_{(0)}^{\rm mx}]$ one has a vanishing trace. Given the structure in eq.~(\ref{mxcU0}) and the result
\begin{align}
\mqInt\,	\Delta \left[\partial_q^{.2},\frac{q_{\alpha}q_{\beta}}{q^2}\right]\Delta=0\,, \label{LoopInt0}
\end{align}
the last term in eq.~(\ref{mxtcU2}) cancels. The first term on the RHS of eq.~(\ref{mxtcU2}) contains the integrals
\begin{align}\label{loopInt1}
	&\mqInt \Delta \left[\partial_q^\nu,\frac{q_\mu}{q^2}\right]\Delta=\frac{g^\nu_\mu}{2}&
	&\mqInt\Delta \left[\partial_q^\mu,\frac{q_{\alpha}q_{\beta}}{q^2}\right]\partial_q^\nu\Delta=\frac1{12}g^{\mu\nu}g_{\alpha\beta}-\frac16 g^{\mu}_{(\alpha }  g^\nu_{\beta)}
\end{align}
so that
\begin{align}
\mqInt \, (i \Prp  [\partial_q^\mu,\mathcal U^{\rm mx}_{(1)}]\Prp)&=\frac i2\left((\psi_{;\nu})^\dagger\moe^\nu g^{\alpha\beta}-\frac{(\psi^{;(\alpha})^\dagger\moe^{\beta)}}{2}\right)\sigma^\mu (g_{\rho\sigma}\moe^.\psi_{;.}+\moe_{(\rho}\psi_{;\sigma)})\\ \nonumber
&+\frac{m_\phi^2\phi}{2}\left(g^{\alpha\beta} g_{(\rho}^\mu\phi_{;\sigma)}+(g^{\nu(\alpha} \phi^{;\beta)}-g^{\alpha\beta}\phi^{;\nu})g_{\nu}^\mu g_{\rho\sigma} \right)\\ \nonumber
&-\left(\frac1{12}g^{\mu\omega}g_{\gamma\nu}-\frac16 g^{\mu}_{(\gamma }  g^\omega_{\nu)}\right)\Bigg(\left(g^{\gamma(\alpha}\phi^{,\beta)}-g^{\alpha\beta}\phi^{,\gamma}\right)g^\nu_{(\rho}\phi_{,\sigma)}\\ \nonumber
&\qquad-\left(g^{[\lambda(\alpha}F^{\beta) \gamma]}-g^{\alpha\beta}F^{\lambda\gamma}\right)\left(g_{[\lambda(\rho} F_{\sigma)\delta]}-g_{\rho\sigma}F_{\lambda\delta}\right)g^{\delta\nu}\Bigg)_{;\omega}
\end{align}
however when tracing the above times $V\propto \varepsilon^{.\alpha\rho.}g^{\beta\sigma}$ all terms but the fermionic one cancel:
\begin{align}\nonumber
&\mqInt  {\rm tr}(\Prp i V[\partial_q,\mathcal U_{(1)}^{\rm mx}])\Delta\\
=&\frac{\kappa^4}{16}\left(10\psi^{\dagger;\mu}\moe^\alpha\psi_{;\mu}\psi^\dagger\moe_\alpha\psi-2\psi^{\dagger ;(\alpha}\moe_{\alpha}\psi^{;\mu)}\psi^\dagger\moe_{\mu}\psi-6i(\psi^{\dagger;\alpha}\moe^\mu\psi^{;\rho})(\psi^\dagger\moe^\nu\psi)\varepsilon_{\alpha\mu\rho\nu}\right)\,.
\end{align}
The remaining term in eq.~(\ref{mxtcU2}) cancels as can be seen as follows introducing the notation $\mathcal U^{\rm mx}{(0)}=\mathcal U^{\rm mx,\mu\nu}_{(0)}q_\mu q_\nu/q^2=\mathcal U^{\rm mx,\star\star}_{(0)}q_\star^2/q^2$
\begin{align}\nonumber
{\rm tr}(\left[V\partial_q,\left[V\partial_q,\mathcal U^{\rm mx,\star\star}_{(0)}\frac{q_\star^2}{q^2}\right]\right])=&{\rm tr}(\left[V\partial_q,V_.\mathcal U^{\rm mx,\star\star}_{(0)}\left[\partial_q^., \frac{q_{\star}^2}{q^2}\right] +\left[V,\mathcal U^{\rm mx,\star\star}_{(0)}\right]\frac{q_\star^2}{q^2}\partial_q^.\right])\\
=&{\rm tr}(V_.V_. \mathcal U^{\rm mx,\star\star}_{(0)} [\partial_q^{.2},\frac{q_\star^2}{q^2}]+V_.\left[V_.,\mathcal U^{\rm mx,\star\star}_{(0)}\right]\left[\partial_q^.,\frac{q_\star^2}{q^2}\right]\partial_q^.)
\end{align}
again given that the integral in eq.~(\ref{LoopInt0}) cancels one has
\begin{align}
\mqInt 
{\rm tr} (\Prp [V\partial_q,[V\partial_q,\mathcal  U^{\rm mx}_{(0)}]]\Prp )=&\left( \frac{1}{12}g^{\mu\nu}g_{\alpha\beta}-\frac16 g^{(\mu}_{\alpha }  g^{\nu)}_{\beta}\right){\rm tr} (V_\mu\left[V_\nu,\mathcal U^{\rm mx,\alpha\beta}_{(0)}\right])\\ \nonumber
=&\left( \frac{1}{12}g^{\mu\nu}g_{\alpha\beta}-\frac16 g^{(\mu}_{\alpha }  g^{\nu)}_{\beta}\right){\rm tr} ([V_\mu,V_\nu],\mathcal U^{\rm mx,\alpha\beta}_{(0)}])=0\,.\label{mxtU2cancels}
\end{align}
The terms in eq.~(\ref{mxtcU2}) then reduce to: 
\begin{align} \nonumber
&\mqInt \Prp\left( \tcU^{\rm mx}_{(2)}-\mathcal U_{(2)}^{\rm mx}\right)\Prp\\
=&\frac{\kappa^4}{16}\left(10\psi^{\dagger;\mu}\moe^\alpha\psi_{;\mu}\psi^\dagger\moe_\alpha\psi-2\psi^{\dagger ;(\alpha}\moe_{\alpha}\psi^{;\mu)}\psi^\dagger\moe_{\mu}\psi-6i(\psi^{\dagger;\alpha}\moe^\mu\psi^{;\rho})(\psi^\dagger\moe^\nu\psi)\varepsilon_{\alpha\mu\rho\nu}\right)\,.
\end{align}
Now we turn to the term $\mathcal U_{(2)}^{\rm mx}$ given in eq.~(\ref{mxcU2}).
Useful relations for the trace of the operator are
\begin{align}
	(g^{\mu(\alpha}\phi^{;\beta) \gamma}-g^{\alpha\beta} \phi^{;\mu \gamma})g^{\nu}_{(\alpha}\phi_{;\beta) \delta}=&2\phi^{;\alpha \gamma} \phi_{;\alpha \delta}g^{\mu\nu}\,,\\
	\left(g^{[\lambda(\alpha}F^{\beta)\mu]}-g^{\alpha\beta}F^{\lambda\mu}\right)^{;\gamma}\left(g_{[\lambda (\alpha}F_{\beta ) \nu]}-g_{\alpha\beta}F_{\lambda\nu}\right)_{;\delta}=&4 F^{\alpha \mu;\gamma}F_{\alpha\nu;\delta}+2g^\mu_\nu F^{\alpha\beta;\gamma}F_{\alpha\beta;\delta}\,,\\
	(q^{(\alpha}\phi^{\beta)}-g^{\alpha\beta}q\phi^;)\frac{1}{q^2}g_{\alpha\beta}\phi_{;.}\partial_q^. +g^{\alpha\beta}\phi_{;.} \partial_q^. \frac{1}{q^2} q_{(\alpha}\phi_{;\beta)}=&2\phi_{;\mu}\phi^{;\nu}\left[\partial_q^\mu\,,\frac{q_{\nu}}{q^2}\right] \,,
\end{align}
and the possible integrals reduce to those in eqs.~(\ref{loopInt1},\ref{LoopInt0}) plus the following
\begin{align}
	\mqInt\Delta\left(\CDv_{(1)}^\mu\frac{ q^{\nu}}{q^2}+ \frac{q^\mu}{q^2}\CDv_{(1)}^{\nu}-\frac{q^\mu}{q^2}\left\{q,\CDv_{(1)}\right\}\frac{q^{\nu}}{q^2}\right)\Delta&=\frac{g^{\mu\nu}R+2R^{\mu\nu}}{24} \,,
\end{align}
so that the result is
\begin{align}  \nonumber
\mqInt \Prp\left(\mathcal U_{(2)}^{\rm mx}\right)\Prp
 \label{U2T}
=&\frac{\kappa^2}{3}\left(F^{\alpha\beta;\lambda}F_{\alpha\beta;\lambda}-2F^{\alpha\beta;\lambda}F_{\alpha\lambda;\beta}-2F^{\alpha\mu}_{\,\,\,\,\,\,\,;\mu}F_{\alpha\nu}^{\,\,\,\,\,\,;\nu}\right)+3m_\phi^2\kappa^2\phi_;^2-4m_\phi^4\kappa^2\phi^2\\
&+\frac {i\kappa^2}2\left((\psi_{;\nu})^\dagger\moe^\nu g^{\alpha\beta}-\frac{(\psi^{;(\alpha})^\dagger\moe^{\beta)}}{2}\right)_{;\mu}\sigma^\mu(g_{\alpha\beta}\moe^\lambda\psi_{;\lambda}+\moe_{(\alpha}\psi_{;\beta)})\\
&+\frac{\kappa^2}2 R\phi_{;.}^2+\frac {\kappa^2}6R  (FF)-\frac{2\kappa^2}3 (FFR)\,,
\end{align}
where $(FFR)=F^{\mu\nu}F_{\nu\rho}R^{\rho\mu}$.

The square of the $\tcU_{(0)}^{\rm mx}$ term involves a trace and a simple integral, carrying on the notation of eq.~(\ref{mxtU2cancels}),
they combine into,
\begin{align}\nonumber\mqInt{\rm tr}\left(\Prp\tcU_{(0)}^{\rm mx}\right)^2\Delta&
	={\rm tr}(\mathcal U^{\rm mx,\mu\nu}_{(0)}\mathcal U^{\rm mx,\lambda\omega}_{(0)})
		 \frac{g_{\mu\nu}g_{\lambda\omega}+g_{\mu(\lambda} g_{\omega)\nu}}{48}\\
&	= 2\kappa^{4}\phi_;^4+\frac{7\kappa^{4}}6 (FF)^2+\frac{4\kappa^{4}}3(FFFF)+3 \kappa^{4}(\phi_; FF\phi_;)\,,
\end{align}
where $(FF)=$tr$FF=F_{\alpha\beta}F^{\beta\alpha}$, $(FFFF)=F^{ab}F_{bc}F^{cd}F_{da}$.

Lastly the crossed term, given the integrals
\begin{align}
	&\mqInt \left\{\Prp ,\Prp\frac{q_\alpha q_\beta}{q^2}\right\}=\frac {g_{\alpha\beta}}4\,,&&\mqInt \left\{\Prp\frac{q_\alpha q_\beta}{q^2} ,\Delta\{q,\CDv_{(1)}\}\right\}=-\frac R6\frac{g_{\alpha\beta}}{4}\,,
\end{align}
results in
\begin{align}\nonumber
	&\mqInt\left\{\Prp \tcU_{(0)}^{\rm mx}\,,\Prp\left(\tcU_{(0)}^{\rm s}-\{q,\tCDv_{(1)}\}\right)\right\}\Prp
=\frac14 {\rm tr}\left(\tcU^{\rm mx,\alpha\beta}_{(0)}g_{\alpha\beta}(\tcU^{\rm s}_{(0)}+R/6)\right)\\ \nonumber
=&\frac{\kappa^2}4\left[\tcU_{(0)}^{\rm s}\right]^{\alpha\beta}_{\rho\sigma}\left(g_{\nu(\alpha}\phi_{;\beta) }(g^{\nu(\rho}\phi^{;\sigma) }-g^{\rho\sigma} \phi^{;\nu })- \left(g_{[\lambda (\alpha}F_{\beta \} \nu]}-g_{\alpha\beta}F_{\lambda\nu}\right)\left(g^{[\lambda(\rho}F^{\sigma)\nu]}-g^{\rho\sigma}F^{\lambda\nu}\right)\right)\\
&	+\frac{\kappa^2}3 R\phi_;^2+\frac{\kappa^2}2 (FF)R\,,
\end{align}
So to summarize, we have that 
\begin{align} \nonumber
&\mqInt\mathcal I_{\rm mx}=\mqInt \left(\Prp \tcU^{\rm mx}_{(2)}+\left\{\Prp\left(\tcU_{(0)}^{\rm s}-\left\{q,\tCDv_{(1)}\right\}\right)\,,\Prp \tcU^{\rm mx}_{(0)}\right\}+\left(\Prp \tcU_{(0)}^{\rm mx}\right)^2 \right)\Prp\\ \nonumber
=&\frac{\kappa^2}4\left[\tcU_{(0)}^{\rm s}\right]^{\alpha\beta}_{\rho\sigma}\left(g_{\nu(\alpha}\phi_{;\beta) }(g^{\nu(\rho}\phi^{;\sigma) }-g^{\rho\sigma} \phi^{;\nu })- \left(g_{[\lambda (\alpha}F_{\beta \} \nu]}-g_{\alpha\beta}F_{\lambda\nu}\right)\left(g^{[\lambda(\rho}F^{\sigma)\nu]}-g^{\rho\sigma}F^{\lambda\nu}\right)\right)
\\ \nonumber
&+\frac{\kappa^4}{16}\left(10\psi^{\dagger;\mu}\moe^\alpha\psi_{;\mu}\psi^\dagger\moe_\alpha\psi-2\psi^{\dagger ;(\alpha}\moe_{\alpha}\psi^{;\mu)}\psi^\dagger\moe_{\mu}\psi-6i(\psi^{\dagger;\alpha}\moe^\mu\psi^{;\rho})(\psi^\dagger\moe^\nu\psi)\varepsilon_{\alpha\mu\rho\nu}\right)\\ \nonumber
&+\frac{\kappa^2}{3}\left(F^{\alpha\beta;.}F_{\alpha\beta;.}-2F^{\alpha\beta;.}F_{\alpha.;\beta}-2F^{\alpha\nu}_{\,\,\,\,\,\,\,;\nu}F_{\alpha.}^{\,\,\,\,\,\,;.}\right)+3m_\phi^2\kappa^2\phi_;^2-4m_\phi^4\kappa^2\phi^2\\ \nonumber
&+\frac {i\kappa^2}2\left((\psi_{;*})^\dagger\moe^* g^{\alpha\beta}-\frac{(\psi^{;(\alpha})^\dagger\moe^{\beta)}}{2}\right)_{;\nu}\sigma^\nu(g_{\alpha\beta}\moe^.\psi_{;.}+\moe_{(\alpha}\psi_{;\beta)})\\ \nonumber
&+2\kappa^{4}\phi_;^4+\frac{7\kappa^{4}}6 (FF)^2+\frac{4\kappa^{4}}3(FFFF)+3 \kappa^{4}(\phi_; FF\phi_;)\\
&+\frac{5\kappa^2}6 R\phi_{;.}^2+\frac {2\kappa^2}3\left(R  (FF)- (FFR)\right)\,, \label{MxRst}
\end{align}
and the dimension of operators generated goes from 2 to 10.

\section{Conclusions}
A novel method for computing loop corrections in gravity was presented based on a covariant derivative expansion.
The generalization for the covariant derivative expansion to gravity was carried out explicitly to 3rd order in inverse loop momenta and employed to compute the one loop UV divergences in Hilbert-Einstein gravity with a cosmological constant $\Lambda$ and spin 0,1/2 and 1 matter. Our results are summarized in eqs.~(\ref{FinAct}-\ref{Mixed},\ref{R2UV},\ref{MxRst}).
While the selected target here was the UV, this technique could be extended to obtain the full one loop action in a universal formula akin to the flat case and in doing so study the model independent properties of gravity on the IR. This extension would require pushing to higher orders in inverse loop momenta in the covariant derivative expansion which stands as a computational challenge. Inflation or the recent interest on low energy consequences of the UV completion of gravity are fields where this technique could be put to use.

\acknowledgments
The author acknowledges fruitful discussions with Enrique Alvarez, Diego Blas, Brian Henning and Hitoshi Murayama. This work was supported by World Premier International Research Center Initiative (WPI Initiative), MEXT, Japan.
\bibliography{CDEforGRToshiokan}
\bibliographystyle{JHEP}
\end{document}